\documentclass[prc,twocolumn,floatfix,showpacs,preprintnumbers,amsmath,amssymb]{revtex4}

\usepackage{graphicx}
\usepackage{dcolumn}    
\usepackage{bm}         
\usepackage{subfigure}
\def \beq {\begin{equation}}
\def \eeq {\end{equation}}
\def \beqa {\begin{eqnarray}}
\def \eeqa {\end{eqnarray}}
\def\d{{\rm d}}

\def\me{m_{\rm e}}

\newcommand{\req}[1]{(\ref{#1})}
\newcommand{\vect}[1]              
           {\mbox{\boldmath$#1$}}  
\begin{document}
%

\title{Theoretical study of elastic electron scattering off stable and
exotic nuclei}

\author{X. Roca-Maza}
\email{roca@ecm.ub.es}
\author{M. Centelles}
\author{F. Salvat}
\author{X. Vi\~nas}

\affiliation{Departament d'Estructura i Constituents de la Mat\`eria
and Institut de Ci\`encies del Cosmos, 
Facultat de F\'{\i}sica, Universitat de Barcelona,
Diagonal {\sl 647}, {\sl 08028} Barcelona, Spain}

%
\begin{abstract}
Results for elastic electron scattering by nuclei, calculated with
charge densities of Skyrme forces and covariant effective Lagrangians
that accurately describe nuclear ground states, are compared against
experiment in stable isotopes. Dirac partial-wave calculations are
performed with an adapted version of the {\sc elsepa} package.
Motivated by the fact that studies of electron scattering off exotic
nuclei are intended in future facilities in the commissioned GSI and
RIKEN upgrades, we survey the theoretical predictions from
neutron-deficient to neutron-rich isotopes in the tin and calcium
isotopic chains. The charge densities of a covariant interaction
that describes the low-energy electromagnetic structure of the
nucleon within the Lagrangian of the theory are
used to this end. The study is
restricted to medium and heavy mass nuclei because the charge
densities are computed in mean field approach.
Since the experimental
analysis of scattering data commonly involves parameterized charge
densities, as a surrogate exercise for the yet unexplored exotic
nuclei, we fit our calculated mean field densities with Helm model
distributions. This procedure turns out to be helpful to study the
neutron-number variation of the scattering observables and allows us
to identify correlations of potential interest among some of these
observables within the isotopic chains.
\end{abstract}


\maketitle

%
\section{Introduction}

Elastic electron-nucleus scattering has been for many years a very
useful tool to investigate the size and shape of stable nuclei
\cite{Hof56,Don75,Don84,Moya86,Sick01}. Electrons interact with nuclei
basically through the electromagnetic force. If the energy of the
electrons is high enough, they become a relatively clean probe to
explore precisely the internal structure of nuclei, insensitive to
strong interaction effects. In particular, the analysis of electron
scattering data provides most valuable information about the charge
distribution in atomic nuclei \cite{Vri87,Fri95,Ang04}.

Developments in accelerator technology and detection techniques
nowadays allow experimentation with nuclei beyond the limits of
$\beta$-stability. The number of nuclei whose masses have been
measured keeps growing \cite{ENAM04} and this tendency is expected to
continue with the use of radioactive isotope beams (RIB)
\cite{Tan95,Gei95,Mue01}. A new generation of electron-RIB colliders
using storage rings is now under construction in RIKEN (Japan)
\cite{Sud01,Kat03} and GSI (Germany) \cite{GSI02,Sim04}. These
facilities will offer unprecedented opportunities to study the
structure of exotic unstable nuclei through electron scattering in the
ELISe experiment at FAIR in Germany \cite{Sim07} and the SCRIT project
in Japan \cite{Sud05,Wak08}. Therefore, the theoretical investigation
of exotic nuclei with models of purported reliability in stable   
isotopes is a timely and challenging problem. Such effort will test
the ability of the established nuclear theory in the domain of exotic
nuclei, and may as well provide valuable references for future
experiments.

In recent literature, several theoretical studies of elastic
electron-nucleus scattering in exotic nuclei have been reported
\cite{Ant05,Sar07,Karat07,Bertu07,Zai04,Zai05,Zai06}. Some of these
works are concerned with analyzing electron scattering in light
nuclei, where exciting exotic phenomena such as the appearence of
halos may take place. That is the case of e.g.\ $^{6,8}$He
\cite{Ant05,Karat07}, $^{11}$Li \cite{Ant05,Bertu07}, $^8$B
\cite{Karat07,Bertu07}, $^{12}$O and $^{28}$S \cite{Zai04} nuclei,
where possible effects on scattering from the occurrence of halos have
been investigated. Light nuclei require a microscopic treatment of the
scattering interaction to properly deal with the underlying shell
model structure \cite{Karat07}. Other works study the variation of the
charge form factors along isotopic \cite{Ant05,Sar07,Zai05} and
isotonic \cite{Zai06} chains of medium and heavy mass nuclei. It has
been found that, when the number of neutrons (protons) in these
isotopic (isotonic) chains increases, the squared modulus of the
charge form factor and the position of its minima show, respectively,
an upward trend and a significant inward shifting in the momentum
transfer. In addition to electron scattering, it is worth mentioning
that proton scattering may be another valuable tool to investigate the
changes in the charge density of the nucleus, especially at its
interior, as one proceeds to the drip lines along isotopic
chains~\cite{Karat04}. 

To investigate the internal structure of nuclear charge densities, the
de Broglie wavelength of the probe has to be of the order of 1 fm.
This means that the energy of the electron beam has to be of the order
of several hundred MeV\@. Therefore, for accurate theoretical
calculations of differential cross sections (DCS) and electric charge
form factors, one needs to solve the elastic scattering of Dirac
particles in the scalar potential pertaining to the nuclear charge
distribution. The simplest approach is the plane-wave Born
approximation (PWBA), where one assumes that the initial and final
states of the electron can be described by plane waves. Although the
PWBA is able to account for important features of scattering, it is
not enough accurate for quantitative calculations of the electric
charge form factor. A more elaborate method is supplied by the Glauber
theory using a relativistic eikonal approximation of the Dirac
equation \cite{Bak64}. It has been successfully applied to a
systematic study in elastic electron-nucleus scattering
\cite{Zai04,Zai05,Zai06}. The most sophisticated calculations of
electron-nucleus scattering employ the exact phase-shift analysis of
the Dirac equation. This method corresponds to the so-called distorted
wave Born approximation (DWBA) \cite{Yen54} and was employed e.g.\ in
Refs.\ \cite{Ant05,Sar07}. In the present work we apply a modified
version of the recently published code {\sc elsepa} \cite{Sal05} to the
elastic electron-nucleus scattering problem. This code was originally
devised to perform accurate Dirac partial-wave calculations of ELastic
Scattering of Electrons and Positrons by Atoms, positive ions, and
molecules in the low-energy domain.

The main input needed for solving the elastic electron-nucleus
scattering problem is the charge density of the target nucleus. In the present
article we use charge densities obtained with mean field models for
calculating electron scattering off medium and heavy 
mass nuclei. For this purpose, we employ effective nuclear
interactions of current use in nuclear structure physics. It is known
that the overall trends of the elastic electron-nucleus scattering in
medium and heavy nuclei are, in general, reasonably reproduced by the
theoretical charge densities obtained in the mean field approximation.
However, different effective interactions predict electric charge form
factors and DCS that differ in fine details and describe with
different quality the experimental data. In our study we consider the
non-relativistic Skyrme forces SkM$^*$ \cite{SkM*} and SLy4
\cite{SLy4}, and the covariant models NL3 \cite{NL3}, FSUGold
\cite{FSUG}, G2 \cite{G2}, and DD-ME2 \cite{Ring}. They are
representative examples of effective interactions that accurately
describe the ground-state properties of finite nuclei and some of
their collective excitations.

The parameters of the alluded microscopic interactions have been
determined from careful calibration to observables such as binding
energies, single-particle levels, and charge and diffraction radii
of a variety of selected nuclei. The nucleon density
distributions of the Skyrme Hartree-Fock and relativistic mean field
theories are obtained by numerical solution of the quantal mean field
variational equations. The effects of the neutrons on the proton
density are taken into account, in a self-consistent manner, through
the interaction terms of the effective force or Lagrangian. Therefore,
no parametrized shapes of the density profiles are implemented.
Nevertheless, most of the mean field calculations of finite nuclei
assume point-nucleon densities. The charge density is obtained from
the proton point-like distribution folded with the proton charge form
factor \cite{Gre92}
\begin{equation}
\rho_p(r) = \frac{\alpha^3}{8 \pi} e^{- \alpha r} .
\label{eq1}
\end{equation}
A value $\alpha^2 = 18.29$ fm$^{-2}$ corresponds to a proton root mean
square (rms) radius of 0.81 fm. This is the standard prescription in
the fitting procedure of the parameters of most Skyrme forces and
relativistic mean field interactions to the experimental data of
finite nuclei, including charge radii. The covariant models of Ref.\
\cite{G2} are an exception to this fact, and in particular the
interaction G2 \cite{G2,Ser97} that we will employ in several of our
calculations. The effective Lagrangian of G2 incorporates the
low-energy electromagnetic structure of the nucleon within the theory
\cite{G2,Ser97}. It is to be emphasized that in G2 the charge density is
obtained directly from the solution of the mean field equations and that there
is no folding to be performed with external single-nucleon form factors,
hereby maximizing the predictive power. In the other mean field forces
considered in our work we will neglect the contribution of the neutron charge
form 
factor \cite{Ber72} to the charge density. This is known to be a
reasonable approximation up to moderate momentum transfers
\cite{Gar99}, which is the region analyzed in the present study of
scattering. Moreover, it ensures consistency with the method
applied originally to fit the parameters of these
interactions; additional modifications into the charge densities
could spoil, e.g., their accurate predictions  for charge radii.

It is to be mentioned that the mean field treatment would reach its
limits in the study of exotic light nuclei, where the shell model
structure and halos can become prevailing features \cite{Karat07}.
Therefore we do not attempt to treat these light systems in the present 
work. The development of suitable tools and a unified framework to
deal with charge densities and electron scattering in nuclei across
the mass table remains an outstanding problem in the field, maybe
appropriate for new initiatives like the UNEDF collaboration to build
a universal nuclear energy density functional \cite{unedf}.

The present article is organized as follows. Section II is devoted to
the theoretical formalism. The mean field description of finite nuclei
in the non-relativistic and relativistic frames is shortly discussed.
The Dirac partial-wave calculation of elastic scattering of electrons
by nuclei implemented in the code {\sc elsepa} is summarized. In Section III,
the elastic electron scattering results obtained from the charge densities of
the above-mentioned effective interactions are studied by comparing with
available experimental data for several stable nuclei. In
Section IV we 
investigate the theoretical predictions for elastic electron-nucleus
scattering in the tin and calcium isotopic chains from the proton drip line to
the neutron drip line. These calculations are performed with the relativistic
mean field interaction G2 \cite{G2}. Our conclusions are laid in the final
section.  

\section{Theory}

In the current section we review the basic features of the Skyrme and
relativistic mean field models that we will employ to compute the
theoretical nuclear charge densities. We also summarize the
calculation of Dirac distorted waves for elastic electron-nucleus
scattering in the code {\sc elsepa}. The reader conversant with
effective nuclear mean field models and with knowledge of the basics
of Dirac partial-wave calculations may prefer to proceed directly to
the discussion of results that starts in Section III.

\subsection{Mean field description of nuclei}

The mean field approach assumes that nucleons move independently
in a mean field generated by the other nucleons of the atomic nucleus.
Useful tools for mean field calculations of nuclei are the
non-relativistic Hartree-Fock method with phenomenological
interactions and the relativistic mean field (RMF) Hartree model with
effective Lagrangians \cite{Ben03}. These phenomenological forces and
effective Lagrangians usually depend on about ten adjustable
parameters that are fitted to reproduce relevant ground-state
properties, such as binding energies and charge radii of a few nuclei.

A common trend of phenomenological interactions used in the mean field
approach is their simple mathematical structure. In the
non-relativistic mean field models, the Skyrme interactions
\cite{Vau72,Li91} are among those most widely used. Skyrme forces are
zero-range interactions, which do not require calculations of
exchange contributions. These forces have been employed for describing
ground-state properties of the atomic nucleus, low-energy excited
states, fission and fusion barriers, nucleon-nucleus and heavy-ion
potentials, etc.\ (see e.g.\ Ref.\ \cite{Li91}). The parameter set
SkM$^*$ \cite{SkM*} is the classical paradigm of such a model. It is
known to yield charge densities in overall agreement with densities
inferred from experiment \cite{Ric03}. SLy4 \cite{SLy4} is a more
modern version of the Skyrme force that was calibrated with special
care for the isospin sector and for predictions of neutron-rich matter
that occurs e.g.\ in neutron stars.

The RMF theory of hadrons has become another useful tool for the study
of bulk and single-particle properties of nuclear matter and finite
nuclei \cite{Ser86,Rei89,Ser92,Ser97}. In the relativistic model,
nucleons are treated as Dirac particles that interact by exchanging
virtual mesons. The covariant theory automatically takes into account
the spin-orbit force, the finite-range, and the density-dependence of
the nuclear interaction. The no-sea approximation, which disregards
effects from the Dirac sea of negative energy states, is adopted. The
open parameters of the model are the meson coupling constants and some
of the meson masses. After fitting them to binding energies, charge
radii, and other well-known empirical data of a few selected nuclei,
the covariant theory predicts average properties of spherical and 
deformed nuclei over the whole periodic table in very good agreement 
with experiment \cite{Gam90,Ring96,G2}.

The original Walecka Lagrangian \cite{Ser86} contained $\sigma$,
$\omega$, and $\rho$ mesons without any meson self-interactions.
It was able to predict the correct saturation point of 
nuclear matter, albeit with a very large incompressibility modulus.
The model was refined with the introduction of sigma meson
self-interactions \cite{Bog77}. A parameterization of this type is
e.g.\ the celebrated NL3 model \cite{NL3}.
These parameter sets properly describe the data about finite
nuclei, but often display differences, at densities above the
saturation point, with microscopic Dirac-Brueckner-Hartree-Fock 
calculations of the nuclear matter equation of state
\cite{Toki94,Est01a}. A better agreement with the latter calculations 
at densities up to 2--3 times the saturation density is achieved by
incorporating a quartic vector meson self-interaction in the effective
Lagrangian. Another interesting addition is a mixed
isoscalar-isovector coupling \cite{Hor01}, which allows one to modulate
the density dependence of the nuclear symmetry energy. The variation of this
coupling leaves the binding energy and proton 
rms of a finite nucleus almost unaltered, but it considerably modifies
the rms radius of the neutron distribution. A representative instance
of this type of model is the FSUGold parameter set \cite{FSUG}. This
set yields an equation of state that is considerably softer than in
NL3, for both symmetric matter and neutron matter. Apart from the
binding energies and charge radii of nuclei, FSUGold delivers a
satisfactory description of several modes of collective excitations 
having different neutron-to-proton ratios.

The Lagrangian density associated with the G2 parameter set is
inspired by effective field theory methods. It contains all couplings
consistent with the underlying QCD symmetries up to the order
considered in the expansion scheme \cite{Ser97,G2}. In contrast to the
majority of mean field models, G2 describes the low-energy electromagnetic
structure of the nucleon within the theory by means of vector-meson dominance
and derivative couplings to the photon, cf.\ Ref.\ \cite{G2} for details. 
As indicated above, this means that no additional
calculations with external nucleon form factors are needed to obtain
the charge density, and that the electromagnetic effects of the
protons and neutrons are included within the low-energy regime in a
unified framework \cite{G2}. The G2 set explains finite nuclei and nuclear
matter with a commendable level of accuracy. It predicts a soft equation of
state both around saturation and at high densities that is consistent with 
recent measurements of kaon production and flow of matter in energetic
heavy-ion collisions as well as with observations of masses and radii
of neutron stars~\cite{Arumugam04}.

Recent formulations of the RMF theory do not introduce mesonic
self-interactions but make the coupling constants of the mesons
density dependent, like in the DD-ME2 parameter set \cite{Ring}. These
models accurately describe the properties of finite nuclei and, in
addition, the associated equation of state of nuclear and neutron
matter at supra-saturation agrees with the trends of microscopic
Dirac-Brueckner-Hartree-Fock calculations that start from the bare
nucleon-nucleon interaction.

Pairing correlations need to be taken into account for the
calculation of open-shell nuclei. We will describe them in both,
non-relativistic  and relativistic frames, through a modified BCS
approach that takes into account the continuum by means of quasi-bound
levels owing to their centrifugal (neutrons) or centrifugal-plus-Coulomb
barriers (protons) \cite{Est01b}. For the Skyrme models used in this
work,  the pairing correlations are introduced
by using a zero-range density-dependent force whose parameters can be
found in Ref.\ \cite{SLy4}. In the case of the covariant NL3, G2, and
FSUGold models, we describe the pairing correlations by means of a
constant matrix element fitted to reproduce the experimental binding
energies of some selected isotopic and isotonic chains \cite{Est01b}.
In the DD-ME2 parameter set a fixed gap is considered, determined from
experimental odd-even mass differences~\cite{Ring}.

\subsection{Description of electron scattering}

The {\sc elsepa} code \cite{Sal05} was originally
designed for the calculation of elastic scattering of electrons and
positrons by atoms, positive ions, and molecules. We have adapted it
to handle high-energy electron scattering by nuclei. {\sc elsepa} computes
the DCS using the conventional relativistic partial-wave method, which
was first formulated by Yennie et al.\ \cite{Yen54}. The projectile
electron is assumed to feel the electrostatic field of the nuclear
charge distribution. The potential energy of an electron at a
distance $r$ from the center of the nucleus is given by
\beq
V(r) =
- 4\pi e \left(
\frac{1}{r} \int_0^r \rho_{\rm ch} (r') \, r'^2 \d r' \right.
+ \left. \int_r^\infty \rho_{\rm ch}(r') \, r' \d r'\right)
\label{cesc.1}
\eeq
where $\rho_{\rm ch}(r)$ denotes the charge density of the nucleus,
considered to be spherically symmetrical. At the energies of interest
for the electron-nucleus problem, the effect of screening by the
orbiting atomic electrons is limited to scattering angles smaller than
1 degree (see, e.g.\ Ref.\ \cite{Sal05}), which are well below the
angular range covered by the electron-nucleus scattering measurements.
Consequently, electron screening is ignored in the present
calculation. Since the nuclear charge density is assumed to vanish
beyond a certain radius $r_{\rm B}$ (i.e.\ the radius of the box where
the nuclear charge distribution is calculated), the potential
\req{cesc.1} is purely Coulombian, $V(r)=-Ze^2/r$, beyond that radius.
Globally it can be regarded as a Coulomb potential with the
short-range distortion arising from the finite size of the nucleus,
i.e., as a modified Coulomb potential.

The DCS for elastic scattering of spin unpolarized electrons is given
by 
\beq
\frac{\d \sigma}{\d \Omega} = |f(\theta)|^2 + |g(\theta)|^2 ,
\label{els.8}\eeq
where
\begin{eqnarray}
f(\theta) = \frac{1}{2 {\rm i} k} \sum_{\ell=0}^\infty &\bigg\{& 
(\ell+1) \left[ \exp\left( 2 {\rm i} \delta_{\kappa=-\ell-1} \right) -
1 \right] \nonumber \\
&+& \ell \left[ \exp \left( 2 {\rm i} \delta_{\kappa=\ell} \right) - 1
\right]
\bigg\} \, P_\ell(\cos\theta) 
\label{els.2}
\end{eqnarray}
and 
\beqa
g(\theta) = \frac{1}{2 {\rm i} k}
\sum_{\ell=0}^\infty &\bigg[& \exp \left( 2 {\rm i}
\delta_{\kappa=\ell}\right) \nonumber \\
&-& \exp\left( 2 {\rm i} \delta_{\kappa=-\ell-1} \right)
\bigg] \, P_\ell^1(\cos\theta)
\label{els.3}\eeqa
are the direct and spin-flip scattering amplitudes, respectively. Here
$k$ denotes the wave number of the projectile electron,
\beq
c \hbar k =  \sqrt{E(E+2\me c^2)},
\label{els.4}\eeq
and the functions $P_\ell(\cos\theta)$ and
$P_\ell^1(\cos\theta)$ are Legendre polynomials and associated Legendre
functions, respectively.
The phase shifts $\delta_\kappa$ represent the behavior of the Dirac
spherical waves at large $r$ distances (see e.g.\ Ref.~\cite{Rose61}).

For modified Coulomb potentials, the spherical solutions of the Dirac
equation are suitably expressed in the form
\beq
\psi_{E\kappa m}({\bf r}) = \frac{1}{r}
\left( \begin{array}{c}
P_{E\kappa}(r) \, \Omega_{\kappa, m} (\hat{\bf r}) \\ [1mm]
{\rm i} Q_{E\kappa}(r) \, \Omega_{-\kappa, m} (\hat{\bf r})
\end{array} \right).
\label{els.5}\eeq
The functions
$\Omega_{\kappa, m} (\hat{\bf r})$ are the spherical spinors, and
the radial functions $P_{E\kappa}(r)$ and $Q_{E\kappa}(r)$ satisfy
the following system of coupled differential equations \cite{Rose61}:
\beqa
\frac{\d P_{E\kappa}}{\d r} &=&
-\frac{\kappa}{r} P_{E\kappa}
+\frac{E-V+2\me c^2}{c\hbar} \, Q_{E\kappa},
\nonumber \\ [4mm]
\frac{\d Q_{E\kappa}}{\d r} &=&
- \frac{E-V}{c\hbar} \, P_{E\kappa}
+ \frac{\kappa}{r} Q_{E\kappa}.
\label{els.6}\eeqa
The relativistic quantum number $\kappa$ is defined as $\kappa =
(\ell-j)(2j+1)$, where $j$ and $\ell$ are the total and orbital angular
momentum quantum numbers. Note that $j$ and $\ell$ are both determined
by the value of $\kappa$; $j=|\kappa|-1/2$, $\ell =
j+\kappa/(2|\kappa|)$. In the numerical calculations, the spherical
waves are normalized so that the upper-component radial function
$P_{E\kappa}(r)$ oscillates asymptotically with unit amplitude.

For modified Coulomb potentials and $r\rightarrow\infty$, we have (see,
e.g., Ref.\ \cite{Sal95})
\beq
P_{E \kappa} (r) \simeq
\sin \left( kr - \ell \frac{\pi}{2} - \eta \ln 2kr + \delta_\kappa
\right),
\label{els.7}\eeq
where
\beq
\eta = Z e^2 \me /(\hbar^2 k)
\eeq
is the Sommerfeld parameter. It is convenient to express the phase shifts
$\delta_\kappa$ as $\Delta_\kappa + \hat\delta_\kappa$, 
where $\Delta_\kappa$ is the phase shift of the point-nucleus Coulomb potential 
and $\hat{\delta}_\kappa$ is the ``inner'' phase shift of the 
short-range potential induced by the nuclear charge distribution.

As indicated above, the calculations reported here have been performed
using the computer code {\sc elsepa} \cite{Sal05}. It solves the
radial Dirac equations using a robust integration algorithm, described
in Refs.\ \cite{Sal95,Sal93}, which effectively minimizes the effect
of truncation errors.
The algorithm starts from a table of values of the function $rV(r)$ at
the points $r_i$ of a radial grid, which is provided by the user. This
function is replaced by the natural cubic spline that interpolates the
tabulated values; thus, in the interval between consecutive grid points,
the potential function $rV(r)$ is represented as a cubic polynomial. The
radial wave equations (\ref{els.6}) are then solved by using the exact 
power-series expansions of the radial functions \cite{Sal93}. The
integration is started at $r=0$ and extended outwards up to a point
$r_{\rm m}$ that is beyond the starting radius $r_{\rm B}$ of the
Coulomb tail. For $r > r_{\rm m}$, the field is purely Coulombian, 
and the normalized upper-component radial Dirac function can be
expressed as 
\beq
P_{E\kappa}(r) = \cos \hat\delta_\kappa \, f_{E\kappa}^{\rm (u)}(r)
+ \sin \hat\delta_\kappa  \, g_{E\kappa}^{\rm (u)}(r),
\label{els.78}\eeq
where $f_{E\kappa}^{\rm (u)}(r)$ and $g_{E\kappa}^{\rm (u)}(r)$
are the upper components of the regular and irregular Dirac-Coulomb
radial functions \cite{Sal95}, respectively.
As usual, the phase shift $\hat\delta_\kappa$ is determined by
matching this outer analytical form to the inner numerical solution at
$r_{\rm m}$, requiring continuity of the radial function
$P_{E\kappa}(r)$ and its derivative. The Dirac-Coulomb functions are
calculated by using the Fortran 
subroutine described in Ref.\ \cite{Sal95}, which delivers values of
the regular and irregular Dirac-Coulomb functions and their
derivatives that are accurate to more than 10 decimal figures.

The convergence rate of the series (\ref{els.2}) and (\ref{els.3}) for
the calculation of $\d \sigma / \d \Omega$ is known to be slow. The
summations are optimized by performing them in two steps \cite{Sal05}.
First, they are evaluated for the pure Coulomb field, for which the
phase shifts $\Delta_\kappa$ are known analytically and the
calculation is fast. Second, the point-nucleus results are subtracted
from the expansions (\ref{els.2}) and (\ref{els.3}); the remaining
series represent the effect of only the short-range component of the
potential and converge more rapidly than the original series. As a
consequence, the number of inner phase shifts $\hat\delta_\kappa$ one
needs to compute is normally much smaller than the number of required
Coulomb phase shifts.

\begin{table*}[t]
\caption{\label{T1} Root mean square radii (in fm) for the studied
charge distributions. 
Exp(fit) values are calculated from the experimental Fourier-Bessel
charge densities \cite{Vri87,Fri95}, except for the case of 
${}^{118}$Sn where a Fermi density is used \cite{Lit72}.}
\begin{tabular}{lcccccccccc} 
 & & & & & & & & \\
\hline\hline 
 & & & & & & & & \\
Nucleus     & &\kern1mm Exp(fit)\kern1mm&DD-ME2&\kern4mm G2\kern5mm &\kern4mm
NL3\kern3mm &FSUGold&\kern3mm SLy4\kern3mm &\kern3mm SkM*\kern2mm \\ 
 & & & & & & & & \\
\hline
 & & & & & & & & \\
${}^{16}$O  & &$2.74 $  &$2.73 $&$2.73 $&$2.73 $&$2.69 $&$ 2.80$&$2.81$\\
${}^{40}$Ca & &$3.45 $  &$3.46 $&$3.46 $&$3.47 $&$ 3.44$&$3.51 $&$3.52 $\\
${}^{48}$Ca & &$3.45 $  &$3.48 $&$3.45 $&$3.47 $&$3.47 $&$3.54 $&$3.54 $ \\
${}^{90}$Zr & &$4.26 $  &$4.28 $&$4.25 $&$4.28 $&$ 4.26$&$4.30 $&$4.29 $ \\
${}^{118}$Sn& &$4.67 $  &$4.63 $&$4.62$&$4.63 $&$4.63 $&$4.65 $&$4.63 $ \\
${}^{208}$Pb& &$5.50 $  &$5.52 $&$5.50 $&$ 5.52 $&$ 5.52$&$5.52 $&$5.51 $ \\
 & & & & & & & & \\
\hline\hline 
\end{tabular}
\end{table*}

In general, {\sc elsepa} gives reliable results for electron energies
up to 1~GeV but numerical difficulties can appear at very small
scattering angles because the DCS for the bare nucleus is very close
to the point nucleus DCS (which is known as the Mott DCS) and
consequently diverges. The code evaluates the finite-nucleus DCS only
for scattering angles larger than $10$ degrees, which is a lower bound
of the usually measured electron-nucleus DCS. Depending on the
particulars of each calculation, difficulties can also be found in the
high-energy regime and for large scattering angles where the DCS takes
much smaller values than the Mott DCS\@. In this case, the nuclear
amplitudes almost cancel the Coulomb scattering amplitudes, thus
magnifying the numerical errors. In practice, round-off errors become
apparent when the nuclear scattering amplitudes are less then
$10^{-5}$ times the Coulomb amplitudes (which are themselves small).
When there are indications that these errors could be important, {\sc
elsepa} discontinues the calculation and the DCS is set to zero.

\section{Elastic electron-nucleus scattering in stable nuclei}

As stated in the Introduction, we compute the charge
density distributions with selected effective nuclear interactions,
namely, the Skyrme forces SkM$^*$ \cite{SkM*} and SLy4 \cite{SLy4},
and the covariant parameterizations NL3 \cite{NL3}, 
FSUGold \cite{FSUG}, G2 \cite{G2}, and DD-ME2 \cite{Ring}. We obtain 
the DCS from these mean field charge densities using the {\sc elsepa} 
code. Specifically, in the present section we will compare the 
theoretical DCS derived from the mean field models
with available experimental data for elastic electron scattering in 
$^{16}$O at 374.5 MeV \cite{Sick70},
$^{40,42,44,48}$Ca and ${}^{48}$Ti at 250 MeV, $^{40,48}$Ca at 500 MeV
\cite{Fro68}, ${}^{90}$Zr at 209.6 and 302 MeV \cite{Phan72},
$^{116,118,124}$Sn at 225 MeV \cite{Lit72}, and $^{208}$Pb at 248.2 and
502 MeV \cite{Fria73}.

In the conventional analyses of the scattering data measured in
experiment, the charge density is usually modeled by means of an
analytical function. For instance, 2- or 3-parameter Fermi
distributions are used to this end. The charge density may also be
constructed from the eigenfunctions of an adjustable single-particle
potential of Woods-Saxon or harmonic oscillator type. Also, nearly
model-independent charge densities are obtained from a Fourier-Bessel
expansion \cite{Dre74} with unknown coefficients. In all these cases
the free parameters in the charge distributions are determined from
the measured electron scattering data through a least-squares
minimization procedure. In some comparisons with our theoretical
predictions we will employ experimental charge densities borrowed from
the literature. These densities have been extracted from fits with
Fourier-Bessel expansions \cite{Vri87,Fri95}. In the case of the
nucleus ${}^{118}$Sn, for which this type of density is not available,
we will employ the 2-parameter Fermi distribution given in Ref.\
\cite{Lit72}. The results computed with these ``experimental'' (fitted)
charge densities will be referred to as Exp(fit) in the tables and
figures henceforth.

First, we report in Table \ref{T1} the rms radii of the theoretical
charge distributions of the nuclei $^{16}$O, $^{40}$Ca, $^{48}$Ca,
$^{90}$Zr, $^{118}$Sn, and $^{208}$Pb predicted by the Skyrme and RMF
forces used in this work. They are compared with the rms radii of the
experimental charge distributions extracted from the analysis of the
electron scattering data \cite{Vri87,Fri95,Lit72}. The agreement is
seen to be very good. This fact is not surprising because some
of the experimental values of the charge radius have been used in the
fit of the free parameters of the Skyrme forces and RMF models 
considered.

\begin{figure}[t]
\begin{center}
\includegraphics[width=0.95\linewidth,angle=0,clip=true]{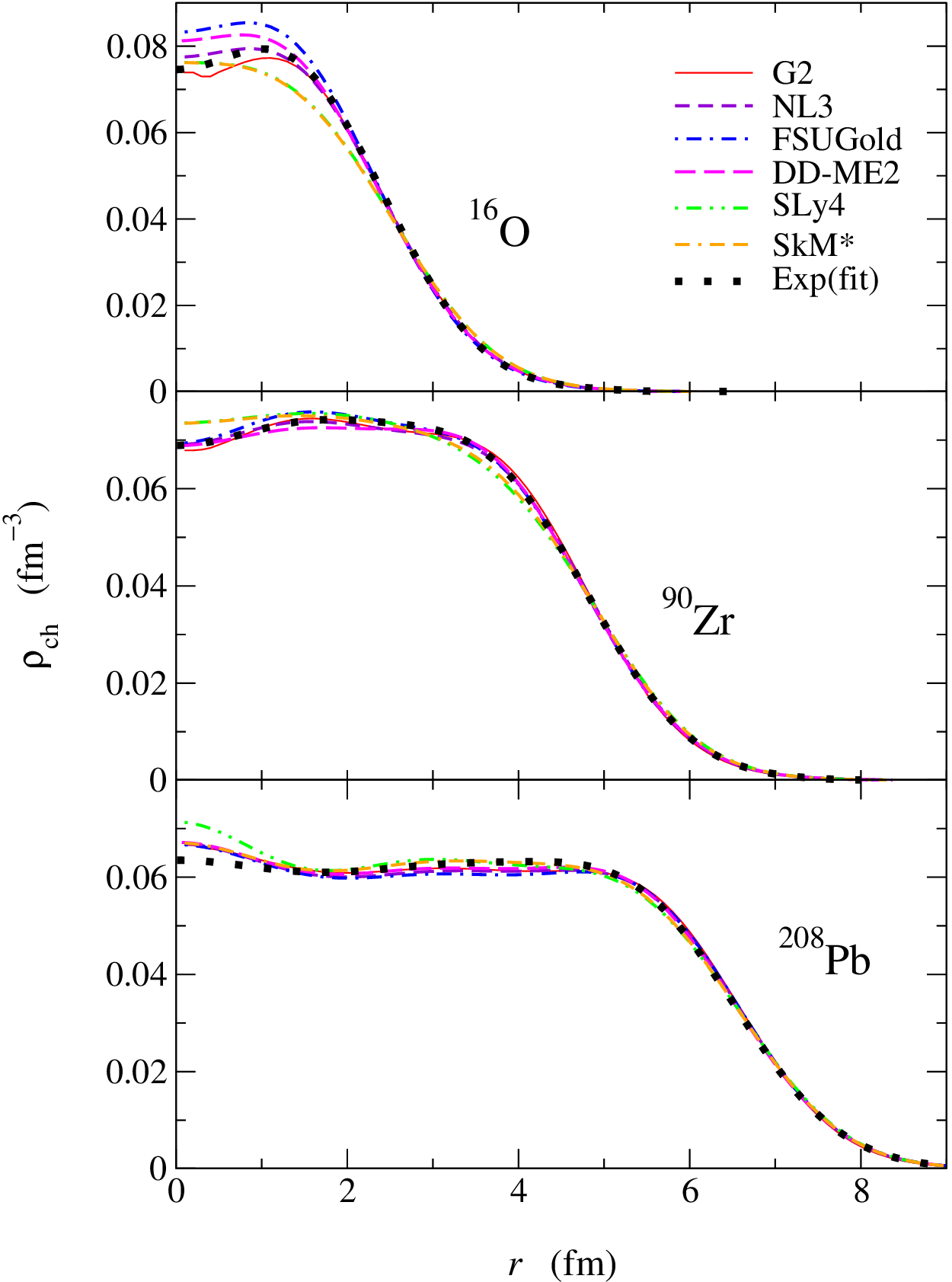}
\caption{\label{F123} (Color online) Radial dependence of the charge
densities of the stable nuclei $^{16}$O, $^{90}$Zr, and $^{208}$Pb. 
The predictions of the different mean field models indicated in the 
legend are compared with the charge densities fitted experimentally 
by the Fourier-Bessel analysis \cite{Vri87,Fri95}.}
\end{center}
\end{figure}

Figure \ref{F123} displays the theoretical charge density profiles obtained
with the investigated mean field models as well as the
experimental charge distributions for $^{16}$O, $^{90}$Zr, and
$^{208}$Pb \cite{Vri87,Fri95}.
In general, theoretical and experimental charge densities agree nicely in 
the fall-off region and differ more in the nuclear interior as a 
consequence of the shell oscillations of the mean field densities.
Differences in the inner region are more marked in light nuclei where a 
mean field approximation may be less justified. In the present case, 
if we analyze the region of the experimental density of $^{16}$O 
between the center and about 2 fm, we see that the covariant sets 
G2 and NL3 give a better description, 
whereas the SkM* and SLy4 forces tend to underestimate 
it and the DD-ME2 and FSUGold sets tend to overestimate it.
A detailed inspection of the 
mean field charge densities shows that they not only differ among themselves
in the nuclear interior but also in the surface region. As discussed in Ref.\ 
\cite{Ric03}, the inner density is normally larger for nuclei with larger 
surface diffuseness. From Fig.\ \ref{F123} we see that for medium and
heavy nuclei the Skyrme charge density, in particular the one predicted by 
SLy4, is larger in the interior and consequently more diffuse at the
surface than the RMF distributions. Differences regarding
the surface diffuseness between non-relativistic and relativistic
charge densities are related to the different density dependence of
the effective interactions \cite{Ric03}.

\begin{figure}[t]
\begin{center}
\includegraphics[width=0.95\linewidth,angle=0,clip=true]{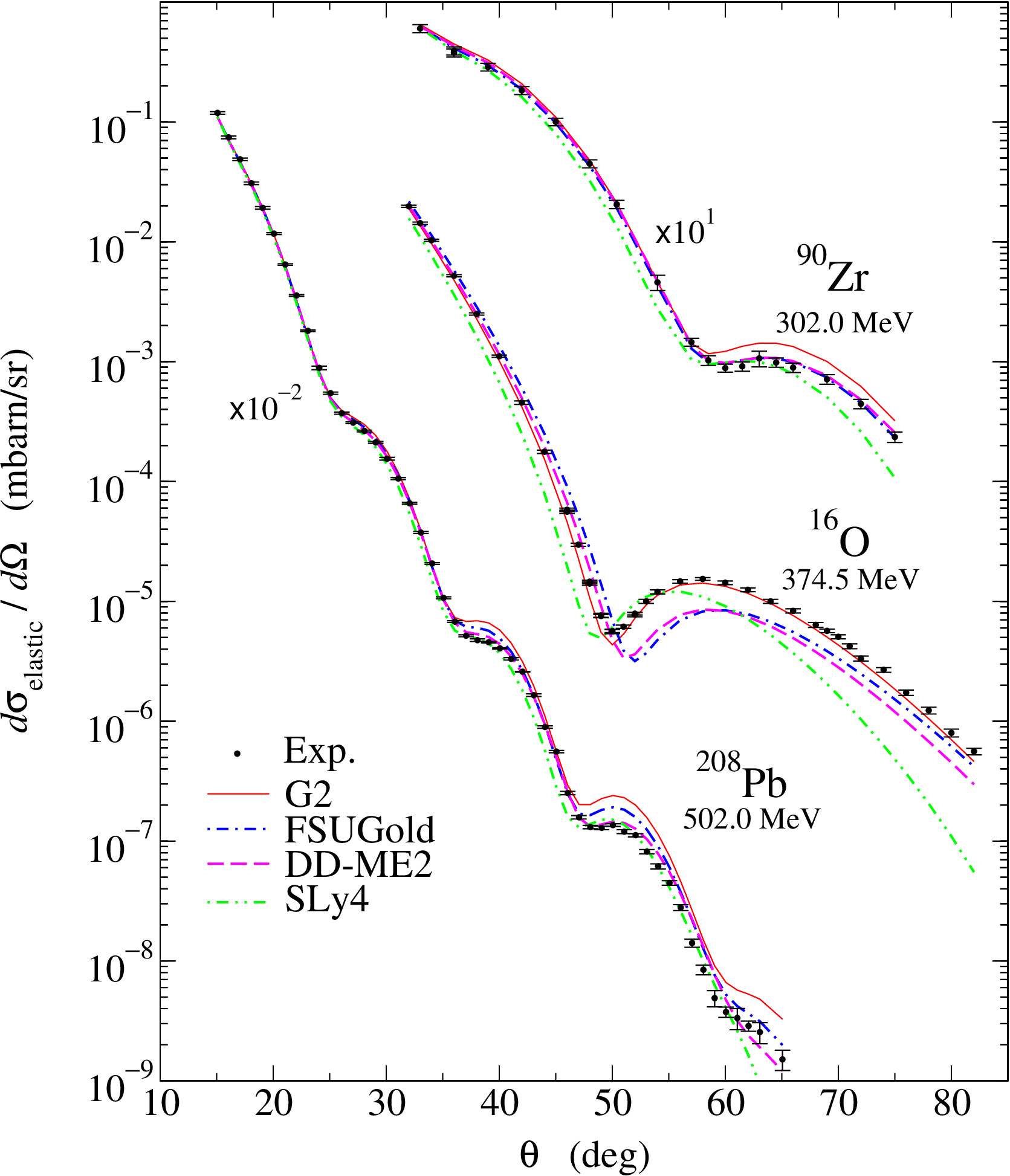}
\caption{\label{F456} (Color online) Elastic DCS for electron-nucleus
scattering in $^{16}$O, $^{90}$Zr, and $^{208}$Pb as a function of the
scattering angle $\theta$, at the energies shown. The results from the 
mean field models indicated in the legend 
are compared with the measured DCS (Exp) \cite{Sick70,Fro68,Phan72} and
with the DCS calculated in the {\sc elsepa} code from the charge densities fitted
experimentally by the Fourier-Bessel analysis [Exp(fit)] \cite{Vri87,Fri95}.
}
\end{center}
\end{figure}

The comparison between theoretical and experimental charge densities
shall be connected with the discussion of the DCS and the electric
charge form factors, which are the quantities measured in real
experiments. Figure \ref{F456} shows the DCS for 
elastic electron scattering in $^{16}$O at 374.5
MeV, $^{90}$Zr at 302 MeV, and $^{208}$Pb at 502 MeV, which we choose
as representative examples to illustrate the predictions of the
different mean field models. 
To improve readability, the curves calculated with NL3
and SkM* are not displayed in the figure.
In the three scattering processes, one 
observes that the experimental data, 
are similarly
reproduced by all of the considered mean field
charge densities at small scattering angles, up to the first
diffraction minimum. The statement is more valid for lead, where models
and data keep close up to the second, or even third, diffraction minimum.
The case which is seen to pose more difficulties to the mean field models
at all scattering angles is, not surprisingly, the lightest investigated 
nucleus, $^{16}$O\@. From the considered interactions, only the parameter set 
G2 is able to reproduce the experimental data of 
$^{16}$O at 374.5 MeV closely at all scattering angles.
As one might expect from the discussion of Fig.\ \ref{F123},
the deviations among the DCS predicted by the various 
mean field models, and the discrepancies with respect to experiment, become 
more prominent at the largest 
scattering angles in the three processes studied in Fig.\ \ref{F456}. 
One detects a significant difference between the non-relativistic
Skyrme interactions and the RMF models used here. 
It is seen
that the SLy4 interaction yields DCS values that, in general, are
smaller than those calculated using the charge densities of
the covariant sets (the same happens with the Skyrme force
SkM*). This trend is especially clear in $^{16}$O and
$^{90}$Zr, if one just excludes the region immediately after the first 
diffraction minimum where a crossing 
of the DCS values predicted by some of the models takes place.

\begin{figure}[t]
\begin{center}
\includegraphics[width=0.95\linewidth,angle=0,clip=true]{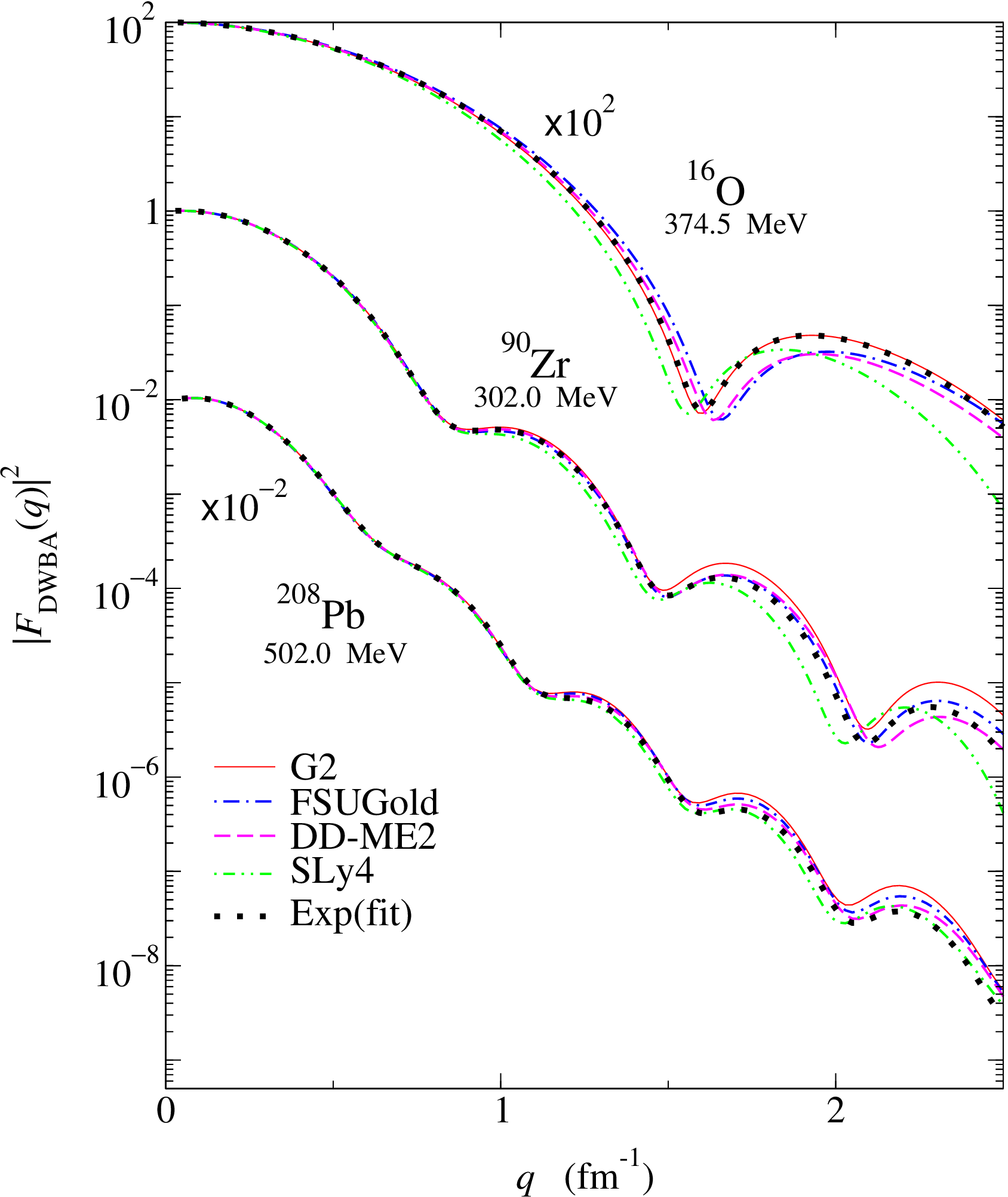}
\caption{\label{F789} (Color online) Squared charge form factor for
$^{16}$O, $^{90}$Zr, and $^{208}$Pb as a function of the momentum transfer
$q$, at the electron beam energies shown. It has been obtained
by applying Eq.\ (\ref{eq12}) as described in the text, 
both for the mean field models indicated in the legend and for the
experimentally fitted charge densities [Exp(fit)] \cite{Vri87,Fri95}.}
\end{center}
\end{figure}

One of the most effective quantities to characterize elastic
electron-nucleus scattering is the electric charge form factor $F(q)$.
The momentum transfer $q$ is related to the scattering angle $\theta$
in the laboratory frame by
\begin{equation}
 c \hbar q = 2 \, E \, \sin(\theta/2) .
\label{qtrans}
\end{equation}
In PWBA the electric-charge form factor is computed as the Fourier
transform of the charge density. In the present work we obtain $|F(q)|^2$ 
at a given beam energy as
\begin{equation}
{\vert F(q) \vert}^2 =
\Big( \frac{d \sigma}{d \Omega} \Big)
\Big( \frac{d \sigma_{\rm M}}{d \Omega} \Big)^{-1} ,
\label{eq12}
\end{equation}
where $d \sigma/d \Omega$ is the DCS calculated from the DWBA analysis
and $d \sigma_{\rm M}/d \Omega$ is the Mott DCS\@. Equation
(\ref{eq12}) goes beyond the first Born approximation to the charge
form factor because, instead of the PWBA for the DCS of the point
nucleus, the exact Mott DCS is used. It will be referred to as $F_{\rm
DWBA}(q)$ in what follows. We will extract the squared charge form factor
from experiment using the same expression (\ref{eq12}), by inserting
the DCS obtained from the experimentally fitted charge densities on
its right-hand side.

\begin{figure*}
\includegraphics[width=0.85\linewidth,angle=0]{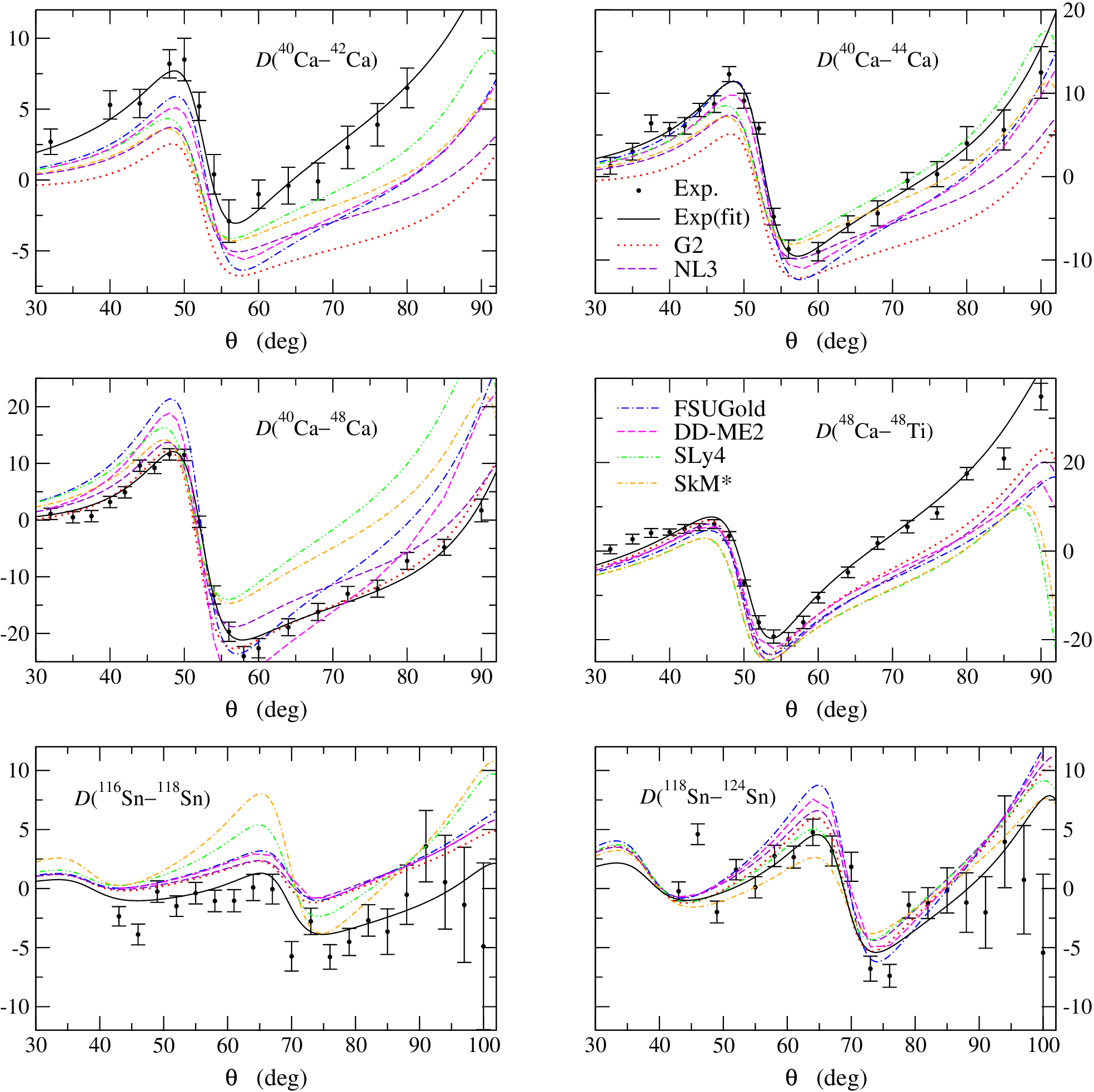}
\caption{\label{F1011121314}(Color online) Relative differences of DCS
in pairs of neighbor nuclei at $250$ MeV for Ca and Ti and at $225$ MeV for
Sn. The results from mean field models are compared with the measured RDDCS
(Exp) and with the RDDCS calculated in the {\sc elsepa} code from the charge
densities fitted experimentally [Exp(fit)] in Refs.\ \cite{Fro68,Lit72}.
Note that in the present figure the vertical scales are linear and
that all of them have been magnified by a factor 100.}
\end{figure*}

Figure \ref{F789} displays the $q$-dependence of ${\vert F_{\rm DWBA}(q)\vert}^2$ 
for the nuclei $^{16}$O, $^{90}$Zr, and $^{208}$Pb at the electron beam energies 
374.5, 302, and 502 MeV, respectively. Results are calculated from the 
considered mean field models and from the experimental charge density 
(as in Fig.\ \ref{F456}, the curves obtained with NL3 and SkM*
are not shown).
As it may be expected from our previous
analysis of the DCS, the experimental values of ${\vert F(q) \vert}^2$
are well reproduced in the low-momentum transfer region by all of the discussed
mean fields. However, some discrepancies appear between the theoretical
predictions and the experimental data at large momentum transfers (or,
equivalently, at large scattering angles). They are an
indication that the theoretical mean field models describe differently
the central region of the experimental charge density \cite{Sick70}, 
which we have addressed in the discussions of Fig.\ \ref{F123}.

\begin{table*}[t!]
\caption{\label{T2} Normalized square weighted difference ($d_{\rm w}^2$)
between the calculated and measured DCS values.
Exp(fit) values are calculated from the experimental Fourier-Bessel
charge densities \cite{Vri87,Fri95}, except for the case of 
${}^{118}$Sn where a Fermi density is used \cite{Lit72}.
The energy $E$ of the incident electrons is in MeV\@.}
\begin{tabular}{lccccccccccc} 
 & & & & & & & & \\
\hline\hline
 & & & & & & & & \\
Nucleus & &$E$ &  &\kern1mm Exp(fit)\kern1mm&DD-ME2&\kern4mm G2\kern5mm
&\kern4mm NL3\kern3mm &FSUGold&\kern3mm SLy4\kern3mm &\kern3mm SkM*\kern2mm \\  
 & & & & & & & & \\
\hline
 & & & & & & & & \\
${}^{16}$O& &$374.5$& &$ 11.1 $&$ 88.7$&$13.1 $&$38.6$&$206.$&$191.$&$194.$\\ 

${}^{40}$Ca& &$250.0$& &$ 7.18 $&$ 3.15$&$16.2$&$13.9$&$0.84$&$24.4$&$24.3$\\ 
           & &$500.0$& &$ 3.48 $&$1.49 $&$42.9$&$19.7$&$5.79 $&$40.0$&$39.0$ \\

${}^{48}$Ca& &$250.0$& &$ 6.66 $&$4.85 $&$9.74$&$7.14$&$4.08$&$14.9$&$13.6$ \\ 
           & &$500.0$& &$ 3.19 $&$1.11 $&$17.0$&$3.53$&$2.57$&$21.84$&$18.5$ \\

${}^{90}$Zr& &$209.6$& &$ 0.78 $&$0.87 $&$2.21$&$1.36 $&$0.65 $&$6.53$&$5.36 $ \\ 
           & &$302.0$& &$ 0.86 $&$0.91 $&$9.92$&$3.27 $&$0.67 $&$9.35$&$7.19 $ \\

${}^{118}$Sn& &$225.0$& &$5.43 $&$18.4 $&$34.8 $&$25.5 $&$31.8 $&$ 2.75$&$ 4.20$ \\ 

${}^{208}$Pb & &$248.2$& &$ 30.6 $&$44.4 $&$154. $&$ 74.8 $&$89.5$&$89.2 $&$61.0 $ \\
            & &$502.0$& &$ 21.2 $&$14.1 $&$186. $&$ 50.5 $&$61.1$&$95.9 $&$76.5 $ \\  
 & & & & & & & & \\
\hline\hline
\end{tabular}
\end{table*}

\begin{table*}[t]
\caption{\label{T3} Normalized square weighted difference ($d_{\rm w}^2$)
  between calculated and measured relative differences of DCS in pairs of 
  neighbor nuclei. The beam energy per electron is 250 MeV for the
  Ca isotopes and $^{48}$Ti, and 225 MeV for the Sn isotopes.}
\begin{tabular}{lcccccccc} 
 & & & & & & & & \\
\hline\hline
 & & & & & & & & \\
$D(A-B)$                     & &\kern1mm Exp(fit)\kern1mm&DD-ME2&\kern4mm
G2\kern5mm &\kern4mm NL3\kern3mm &FSUGold&\kern3mm SLy4\kern3mm &\kern3mm
SkM*\kern2mm \\  
 & & & & & & & & \\
\hline
 & & & & & & & & \\
$D({}^{40}$Ca$-{}^{42}$Ca$)$  & &$0.56 $&$ 9.1$&$ 28.3$&$16.0$&$11.1$&$9.11$&$12.9$\\ 
$D({}^{40}$Ca$-{}^{44}$Ca$)$  & &$1.14 $&$4.5 $&$ 29.6$&$12.2$&$3.88$&$7.08$&$9.13$\\
$D({}^{40}$Ca$-{}^{48}$Ca$)$  & &$1.06 $&$16.4$&$ 4.89$&$7.74$&$38.5$&$94.1$&$49.3$\\
$D({}^{48}$Ca$-{}^{48}$Ti$)$  & &$2.49 $&$18.0$&$ 19.6$&$31.0$&$37.8$&$71.8$&$64.9$\\ 
$D({}^{116}$Sn$-{}^{118}$Sn$)$ & &$2.05 $&$8.05 $&$ 7.80$&$9.00$&$10.1$&$13.2$&$18.5$\\ 
$D({}^{118}$Sn$-{}^{124}$Sn$)$ & &$4.03 $&$5.35 $&$ 6.98$&$7.50$&$9.22$&$7.05$&$7.18$\\ 
 & & & & & & & & \\
\hline\hline
\end{tabular}
\end{table*}

A more exigent test of the mean field charge densities, than the DCS
themselves, is provided by the analysis of the relative differences of
DCS values between pairs of neighbor nuclei (RDDCS). The RDDCS are
defined as
\begin{equation}
D(A-B) = \frac{(d \sigma / d\Omega)_A - (d \sigma / d\Omega)_B}
{(d \sigma / d\Omega)_A + (d \sigma / d\Omega)_B}.
\label{eq11} 
\end{equation}
We explore the following cases where experimental values are
available: $^{40}$Ca-$^{42}$Ca, $^{40}$Ca-$^{44}$Ca, 
$^{40}$Ca-$^{48}$Ca, and $^{48}$Ca-$^{48}$Ti \cite{Fro68}, as well as
$^{116}$Sn-$^{118}$Sn and $^{118}$Sn-$^{124}$Sn \cite{Lit72}.
The theoretical predictions for the RDDCS (\ref{eq11}) from the mean
field charge densities are displayed against experiment in
Fig.\ \ref{F1011121314}, where a factor of 100 has been introduced for
the sake of clarity. Notice that the vertical scale of this figure is
linear instead of logarithmic.
In general, all the considered models describe fairly well the
experimental RDDCS values for small scattering angles. The agreement
with experiment deteriorates when the scattering angle increases,
pointing out again some possible deficiencies in the inner region of
the theoretical charge density distributions.

To make a more quantitative analysis of the electron scattering DCS derived 
from the mean field charge densities, we introduce a normalized square 
weighted difference ($d_{\rm w}^2$), or comparison magnitude, with respect
to the DCS measured in experiment. For each nucleus and electron beam energy,  
it is defined as
\begin{equation}
d_{\rm w}^2 = \frac{1}{N} \sum_{i=1}^N \bigg(
\frac{(d \sigma / d\Omega)^{\rm exp}_i
- (d \sigma / d\Omega)^{\rm calc}_i}
{\delta (d \sigma / d\Omega)^{\rm exp}_i} \bigg)^2 .
\label{eq13}
\end{equation} 
In this expression, the quantities $(d \sigma / d\Omega)^{\rm calc}_i$, $(d
\sigma / d\Omega)^{\rm exp}_i$ and $\delta(d \sigma / d\Omega)^{\rm exp}_i$
are, respectively, the calculated DCS, the measured DCS and the 
uncertainty of the latter. The sum in (\ref{eq13}) runs over all the 
$N$ available data for the given scattering process.

In Table \ref{T2} we report, for several elastic electron-nucleus
reactions, the $d_{\rm w}^2$ values (\ref{eq13}) from theory in comparison with
the $d_{\rm w}^2$ obtained from the charge densities fitted experimentally. One
realizes that the $d_{\rm w}^2$ of the DCS computed with the theoretical charge
densities take sizably varying values, depending on the scattering
process and nuclear model. Nevertheless, for all the considered electron
scattering processes, there are Skyrme forces or RMF parameterizations
that yield a $d_{\rm w}^2$ of similar quality to the experimental charge
density. In particular, the RMF sets DD-ME2 and FSUGold give, on the
average, the best overall agreement with the considered experimental data. 
The Skyrme forces SkM$^*$ and SLy4 provide a similar description, but
globally this description is slightly worse than in the case of the
RMF parameterizations, except for the nucleus ${}^{118}$Sn. 

We have also analyzed the $d_{\rm w}^2$ values of the relative
differences between differential cross sections (\ref{eq11}) in pairs
of neighbor nuclei. The corresponding results are collected in Table
\ref{T3}. For all the mean fields considered in the present study, the
predicted $d_{\rm w}^2$ values are larger than the $d_{\rm w}^2$
values computed with the charge densities that are fitted to
experiment. Looking globally at the results presented in Table
\ref{T3}, the agreement obtained between theoretical and experimental
RDDCS is not significantly worse than the agreement found in the case
of the DCS analyzed previously in Table \ref{T2}. But, in the DCS
analysis, it has been seen in Table \ref{T2} that for each scattering
process there is always one or various specific interactions that are
able to come very close to experiment, whereas in Table \ref{T3} this
does not happen in any of the studied RDDCS in pairs of neighbor nuclei. 

\section{Electron scattering along isotopic chains}

Different tendencies of the square of the charge form factor as a
function of the momentum transfer have been studied in earlier
literature for isotopic \cite{Ant05,Sar07,Zai05} and isotonic
\cite{Zai06} chains of medium and heavy nuclei. Our aim here is to
perform an analysis, quantitative whenever possible, which could
eventually be useful for future electron scattering measurements in
RIB facilities.

We will be concerned with the study of electron scattering in isotopic
chains of medium and relatively heavy systems, which will be
exemplified by the cases of the Ca and Sn isotopes. Many of these
nuclei lie in the region of the nuclear chart that is likely to be
explored employing RIB facilities. Some of them may be investigated in
future electron scattering experiments, such as ELISe \cite{Sim07} and
SCRIT \cite{Sud05,Wak08}. 
For the purpose of our study, we are interested in predicting
the trends of the variation along the isotopic chains of
the electric charge form factor in the low-momentum transfer regime.
Our calculated mean field charge densities will
be parameterized by means of the Helm model \cite{Helm56}, often used
in the analysis of experimental data, with a view to gain deeper
physical insights and to elucidate possible correlations among
scattering observables within the isotopic chains.

We have seen in the previous section that in the region of small $q$
values the experimental results are almost equally well reproduced by
the calculations with the different theoretical mean field models
considered. In our subsequent study of electron scattering in the tin
and calcium chains, as a representative reference, we will work with
the charge densities predicted by the covariant interaction G2
\cite{Ser97,G2}. This model was constructed as an effective hadronic
Lagrangian consistent with the symmetries of quantum chromodynamics.
As mentioned, the model describes the low-energy electromagnetic
structure of the nucleon using vector-meson dominance and provides
directly the charge density of the nucleus, so that no external
single-nucleon form factors are required to compute the latter
\cite{G2}. G2 is also a reliable parameter set for 
calculations of ground-states of nuclei and, at the same time, for
predictions of the nuclear equation of state up to supra-normal
densities and of some properties of neutron stars
\cite{Ser97,G2,Arumugam04}. Calculations of the squared charge form
factor done in PWBA with the set G2 for stable isotopes have been
reported elsewhere \cite{G2}.

\begin{figure}[t]
\begin{center}
\includegraphics[width=0.95\linewidth,angle=0,clip=true]{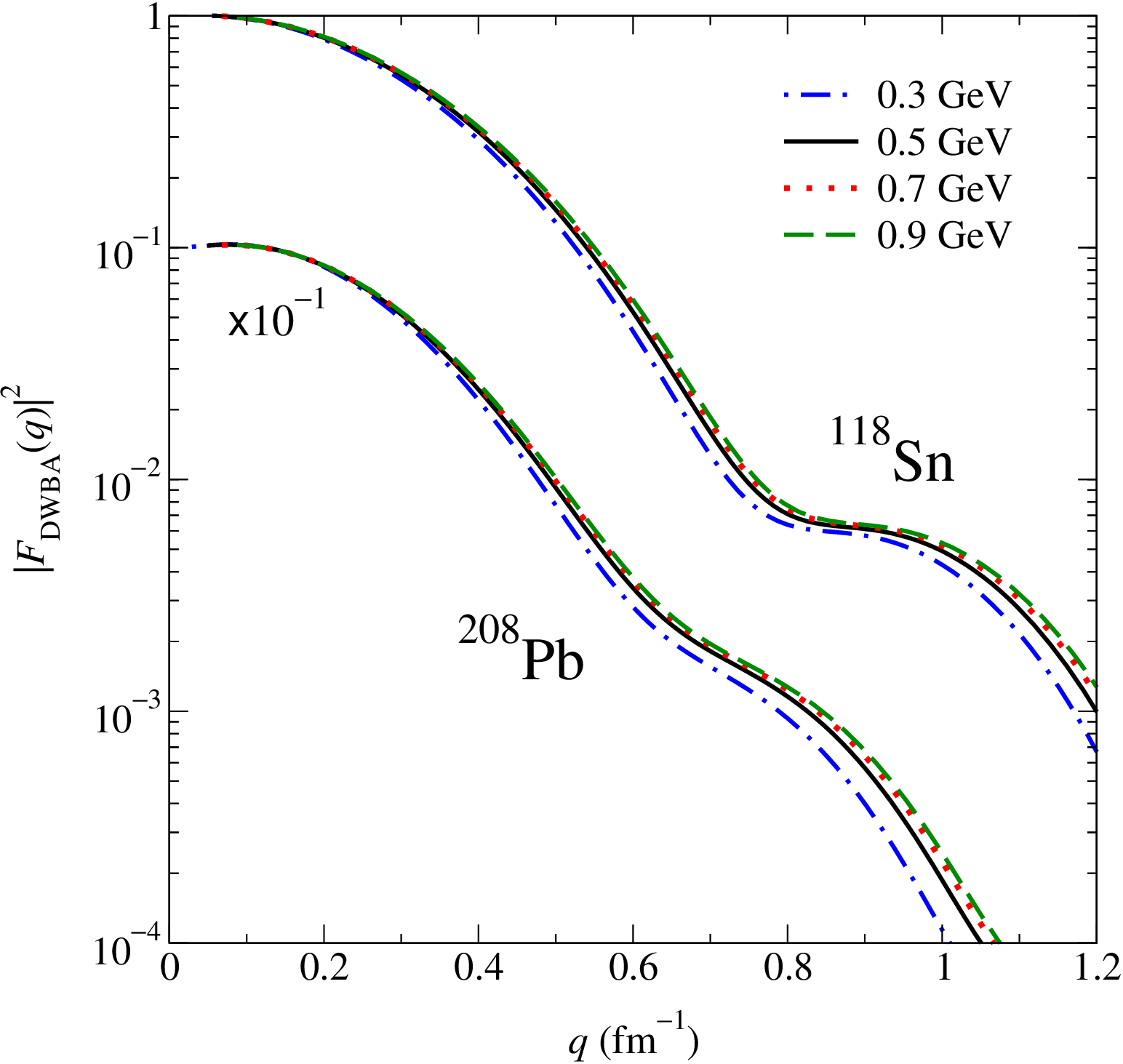}
\caption{\label{ffPbSn} (Color online) 
Squared charge form factor of $^{118}$Sn and $^{208}$Pb, derived from
the covariant mean field model G2, for the indicated electron beam
energies. The figure points out that the sensitivity of the
charge form factor extracted through Eq.\ (\ref{eq12}) to the beam
energy in high-energy elastic electron-nucleus scattering is rather
small, even when the atomic number of the target is large as in
$^{208}$Pb.}
\end{center}
\end{figure} 

For each nucleus in an isotopic chain we compute the associated DCS
via the DWBA calculation using the G2 charge density. The electron
beam energy in all the investigated scattering processes is fixed
at 500 MeV\@. In practice, the energy dependence of the electric
charge form factor defined in Eq.\ (\ref{eq12}) is seen to be
considerably weak for low-momentum transfers, the regime addressed in
our analysis. We have verified numerically that $\vert F_{\rm DWBA}(q)
\vert^2$ of high-energy electron scattering depends little on the
electron beam energy for momentum transfers up to 1--1.5 fm$^{-1}$
(the precise value depends on the nucleus). This happens even for a
heavy system like $^{208}$Pb, where the departure from the
point-nucleus assumption of the Mott DCS used on the r.h.s.\ of Eq.\
(\ref{eq12}) is more significant. We illustrate the situation in Fig.\
\ref{ffPbSn}, where we display $\vert F_{\rm DWBA}(q) \vert^2$ at
several beam energies for electron scattering off $^{118}$Sn and
$^{208}$Pb.

Before proceeding to the presentation of the results in the tin and
calcium isotopic chains, we briefly summarize in the next subsection
how we determine the parameters of the Helm model charge density
distributions.

\subsection{Equivalent Helm charge densities}

Valuable insights into the study of electron scattering often stem
from consideration of modeled charge densities and electric charge
form factors. Moreover, these parameterized forms are instrumental in
experimental data analyses. A notable case is the so-called Helm
model, whose original version \cite{Helm56} has later been extended in
various ways for more accurate descriptions of the experimental charge
densities \cite{Frie82,Frie86,Spru92}. In the simpler version of the
model, two chief features of the nuclear charge density, namely, 
the position and the thickness of the surface, can be related explicitly 
to the electric charge form factor obtained in PWBA\@.
The Helm charge density is obtained from the convolution of a constant
density $\rho_0$ in a hard sphere of radius $R_0$ (the diffraction radius)
with a Gaussian distribution of variance $\sigma^2$ (whose square root 
relates to the nuclear surface thickness):
\begin{equation}
\rho^{(H)}(\vec{r}) = \int d \vec{r}' f_G (\vec{r} -\vec{r}') \rho_0 
\Theta(R_0 - r),
\label{eq14}
\end{equation} 
where 
\begin{equation}
f_G(r) = \big( 2 \pi \sigma^2 \big)^{-3/2} e^{-r^2/2 \sigma^2}.
\label{eq15}
\end{equation}
The rms radius of the Helm density is readily obtained
from Eqs.\ (\ref{eq14}) and
(\ref{eq15}). It can be expressed in terms of $R_0$ and $\sigma$ as
\begin{equation}
\langle r^2 \rangle_H^{1/2} = \sqrt{\frac{3}{5}(R_0^2 + 5 \sigma^2)}.
\label{eq16}
\end{equation}
The corresponding electric charge form factor in PWBA is given by
\begin{equation}
F^{(H)}(q) = \int e^{i \vec{q}\cdot\vec{r}} \rho^{(H)}(\vec{r})
d\vec{r} = \frac{3}{q R_0} j_1(q R_0) e^{- \sigma^2 q^2/2} ,
\label{eq17}
\end{equation}
where $j_1 (x)$ is a spherical Bessel function.

The diffraction radius $R_0$ of the Helm density is usually fixed as
follows. One requires that the first zero of Eq.\ (\ref{eq17}) occurs 
at $q_{\rm min}R_0$, where $q_{\rm min}$ corresponds to the first minimum 
of the modulus of the PWBA form factor ($F_{\rm PWBA}(q)$ hereinafter) 
associated to the original charge distribution that the Helm density 
attempts to describe:
\begin{equation}
R_0 = \frac{4.49341}{q_{\rm min}}.
\label{eq18}
\end{equation}
The variance $\sigma^2$ of the Gaussian is chosen to reproduce the
height of the second maximum of $|F_{\rm PWBA}(q)|$, located at
$q_{\rm max}$: 
\begin{equation}
\sigma^2 = \frac{2}{q_{\rm max}^2} 
\ln \bigg( \frac{3 j_1(q_{\rm max}R_0)}{
q_{\rm max}R_0 F_{\rm PWBA}(q_{\rm max})} \bigg) .
\label{eq19}
\end{equation}

For moderate values of $q R_0$, the Helm charge form factor
$F^{(H)}(q)$ reproduces well the actual charge form factor, with the
exception of the regions closest to its zeros \cite{Helm56}. The
relative difference between $F^{(H)}(q)$ and the actual charge form
factor becomes progressively manifest as the momentum transfer grows.
The applicability of the Helm model near the drip lines is to be
explored. The extent to which it may be appropriate away from
stability, for the purposes of our study, will be validated later from
the numerical point of view in connection with the discussion of the
calculations in the Sn and Ca chains.

\subsection{Tin and calcium isotopic chains}

We are interested in the study of elastic electron scattering in the
Ca and Sn chains. The calculated DCS and squared charge form factors 
$\vert F_{\rm DWBA}(q) \vert^2$ show, for the lightest nuclei considered 
here (calcium isotopes), a relatively well marked first minimum. 
This first minimum, 
however, practically disappears for the heavier nuclei analyzed (tin
isotopes). (This fact can also be told from the previous figures
\ref{F456} and \ref{F789} for $^{16}$O, $^{90}$Zr, and $^{208}$Pb.)
In the latter case, the form factor $\vert F_{\rm DWBA}(q)
\vert^2$ of Eq.\ (\ref{eq12}) still shows an inflection point (IP) in 
the low-momentum transfer region, a point where the curvature changes 
sign. In the absence of an explicit minimum at low-momentum transfer, 
these IP are the best candidates to characterize along the 
isotopic chain relevant properties of the electric charge form factor
in the small-$q$ region.

We determine for each isotope an equivalent Helm distribution from the
calculated mean field charge density. In the PWBA the distortion of
the electron wave functions due to the Coulomb potential of the
nucleus is neglected. The effect of the Coulomb attraction felt by the
electrons can be simulated by replacing the momentum transfer $q$ by
an effective value~\cite{Helm56}
\begin{equation}
q_{\rm eff} = q\bigg( 1 + c \frac{Z \alpha}{q R_{\rm ch}}\bigg),
\label{eq20}
\end{equation}
where $R_{\rm ch}=\sqrt{3/5}R$ is the rms of the charge density
assuming a hard sphere distribution of radius $R = r_0 A^{1/3}$. In
Ref.\ \cite{Helm56} the value of the constant $c$ is taken to be 3/2. 
Here we leave $c$ as a free parameter. It is optimized so that the rms radii
of the equivalent Helm charge densities (\ref{eq16}), with $R_0$ and
$\sigma$ determined by Eqs.\ (\ref{eq18}) and (\ref{eq19}), and with
$q$ replaced by $q_{\rm eff}$, best reproduce along the isotopic chain
the mean field rms charge radii obtained with the RMF parameterization
G2. Proceeding in this way, we find $c\approx 0.15$ for the tin isotopes 
and $c\approx 0.12$ for the calcium isotopes.

\begin{figure}
\begin{center}
\includegraphics[width=0.95\linewidth,angle=0]{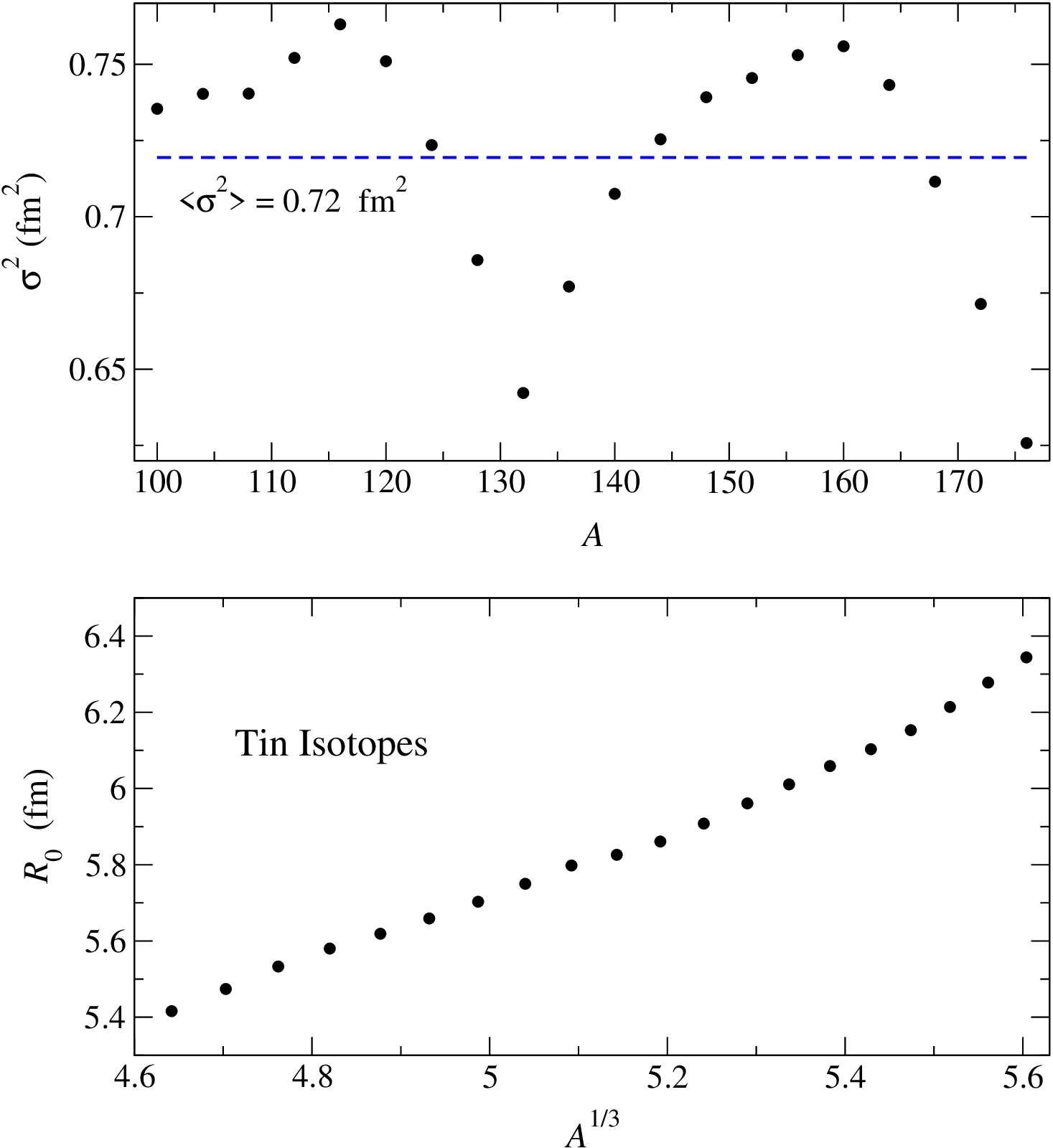}   
\caption{\label{hpartin} Mass-number dependence of the Helm parameters
$\sigma^2$ (upper panel) and $R_{0}$ (lower panel) predicted by the
covariant mean field model G2 in the Sn isotopic chain. The average value of
$\sigma^2$ is depicted by a horizontal dashed line.} 
\end{center}
\end{figure}

We first investigate the tin isotopic chain. 
The calculated Helm parameters $R_0$ and $\sigma^2$ 
are displayed in the lower and upper panels of Figure
\ref{hpartin}, respectively. It is seen that $R_0$ steadily increases
with the mass number $A$ and that it roughly follows the typical
$A^{1/3}$ law. On the contrary, the change of the variance $\sigma^2$
with mass number shows a non-uniform character along the
isotopic chain, related to the underlying shell structure.
It oscillates around a mean value $\sigma^2 \simeq 0.72$
($\sigma \simeq 0.85$).
It is to be noted that $\sigma^2$ displays local minima at the
doubly magic isotopes $^{132}$Sn and $^{176}$Sn (neutron drip line
nucleus). This fact points out a stiffer nuclear surface for
neutron-rich nuclei with double-closed major shells, in agreement with
earlier literature~\cite{Frie82}.

\begin{figure}[t]
\begin{center}
\includegraphics[width=0.98\linewidth,angle=0]{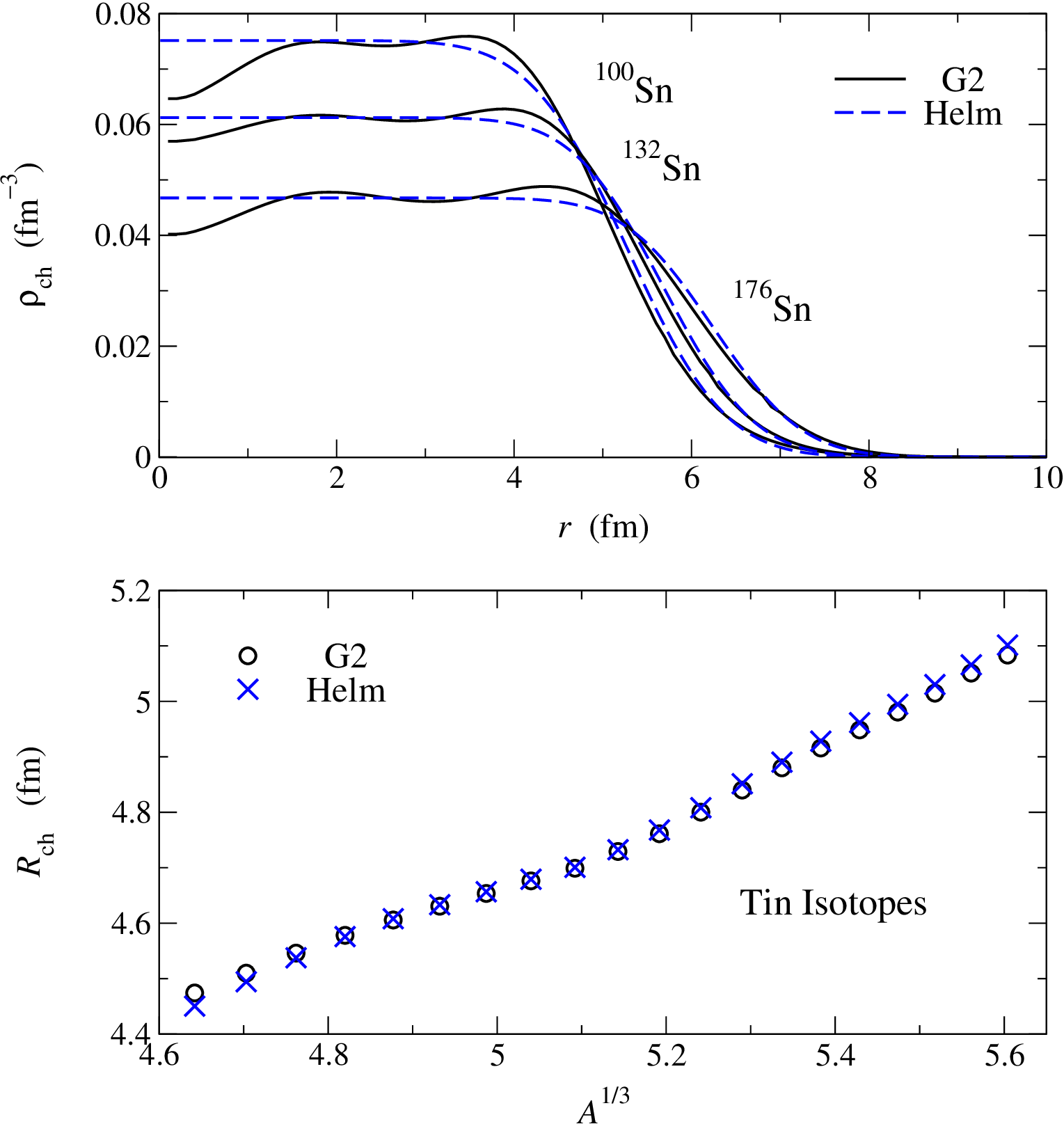}
\caption{\label{drchtin} Charge densities (upper panel) and charge radii
  (lower panel) in the Sn isotopic chain, according to the covariant model 
G2 and to the determined Helm distributions.} 
\end{center}
\end{figure}

The upper panel of Figure \ref{drchtin} depicts the radial dependence
of the mean field charge density profiles computed with the G2
covariant interaction for the isotopes $^{100}$Sn, $^{132}$Sn, and
$^{176}$Sn. We have selected these examples to illustrate the
evolution of the results along the isotopic chain from one drip line
to another. Though the three isotopes share the same
atomic number, one notices outstanding variations in the calculated
mean field charge densities of these isotopes, both in the surface
region and, especially, in the interior region. This fact reflects the
important influence of the changing neutron number on driving the
structure of the charge density along the isotopic chain. This
influence is encoded in the interaction terms of the covariant
Lagrangian of G2 through the exchanged mesons and the couplings to the
photon, and it is accounted for self-consistently by the mean field
calculations. In the same figure we compare the fitted Helm charge
densities and the original G2 charge densities, for the three
discussed tin isotopes. At the interior of these nuclei, we see that
the uniform density of the Helm model averages the oscillations of the
self-consistent quantal densities obtained with the G2 interaction. In spite
of the fact that the surface fall-off of the Helm densities is of Gaussian
type, the agreement at the surface region between the mean field charge 
densities and their Helm equivalents is fairly good as one
proceeds along the isotopic chain, from stability to the neutron and
proton drip lines. This provides some confidence on using the Helm
model as determined in the present work
when the drip lines are approached in the Sn chain.
The lower panel of Figure \ref{drchtin} illustrates the mass number
dependence of the mean field and the equivalent Helm rms charge radii.
Our approach leads to an excellent agreement between the Helm
values and the mean field values in the Sn chain. One appreciates
some slight discrepancies only for isotopes very close to the drip
lines. 
The rms charge radii of both calculations
follow the expected linear trend with $A^{1/3}$. One observes some
departure from the $A^{1/3}$ behavior to slightly lower values
in the isotopes close to $A=132$.
This is finally the reason why the variance $\sigma^2$ decreases
around $A=132$ in Figure \ref{hpartin} [also see Eq.\ (\ref{eq16})].

\begin{figure}
\begin{center}
\includegraphics[width=0.95\linewidth,angle=0,clip=true]{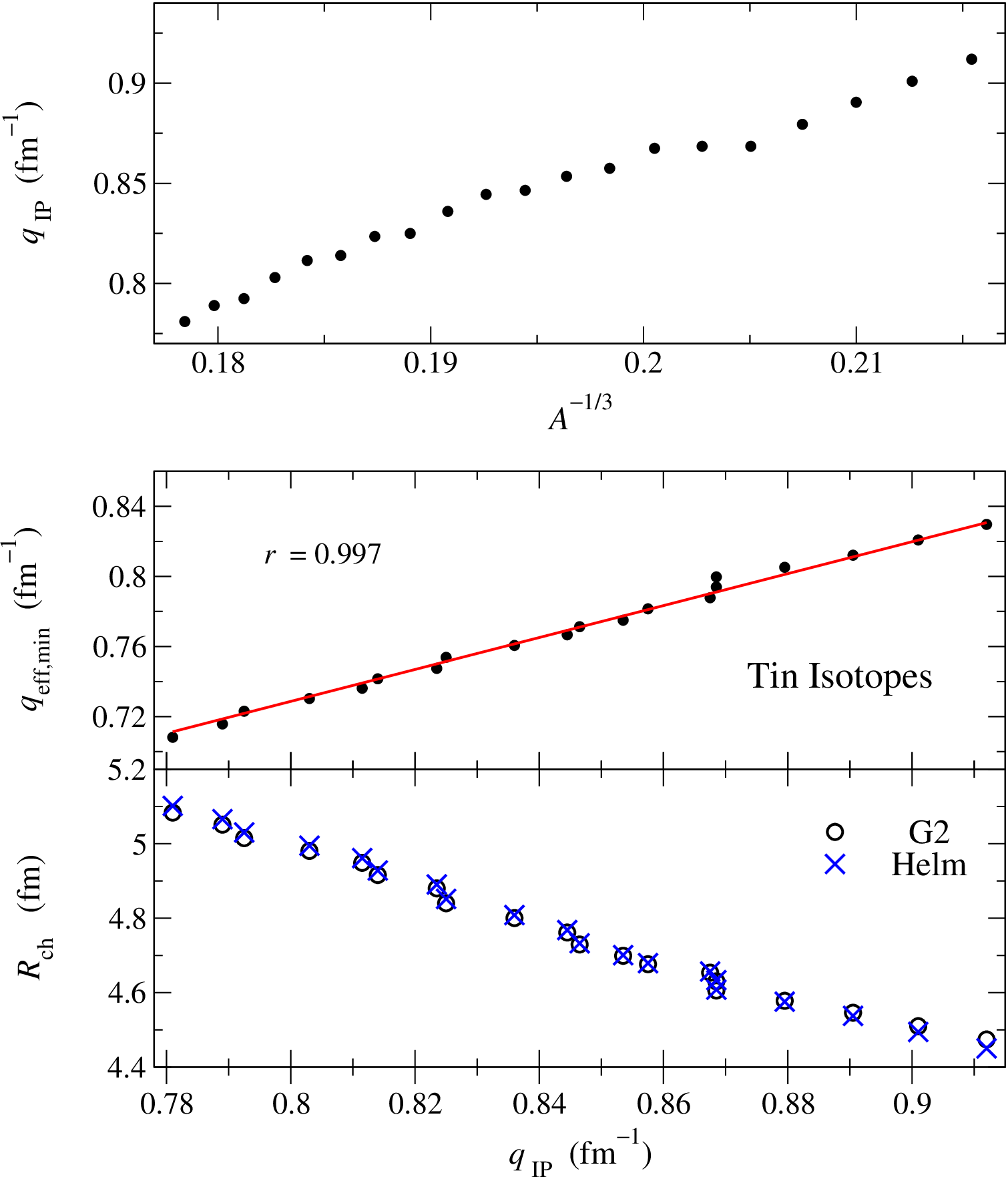}
\caption{\label{qartin} Results predicted by the G2 effective
interaction for 500 MeV electron scattering in the Sn isotopic chain.
Upper panel: mass-number dependence of the momentum transfer at the
first inflection point ($q_{\rm IP}$) of the squared charge form factor
in DWBA. Middle panel: correlation of the effective momentum transfer
at the first minimum of the squared charge form factor in PWBA
($q_{\rm eff,min}$) with the value of $q_{\rm IP}$. A linear fit of
the results is shown and the correlation coefficient $r$ is indicated.
Lower panel: the change of the charge radii calculated with G2 and
with the corresponding Helm densities is depicted against the
value of $q_{\rm IP}$.}
\end{center}
\end{figure} 

\begin{figure}
\begin{center}
\includegraphics[width=0.95\linewidth,angle=0,clip=true]{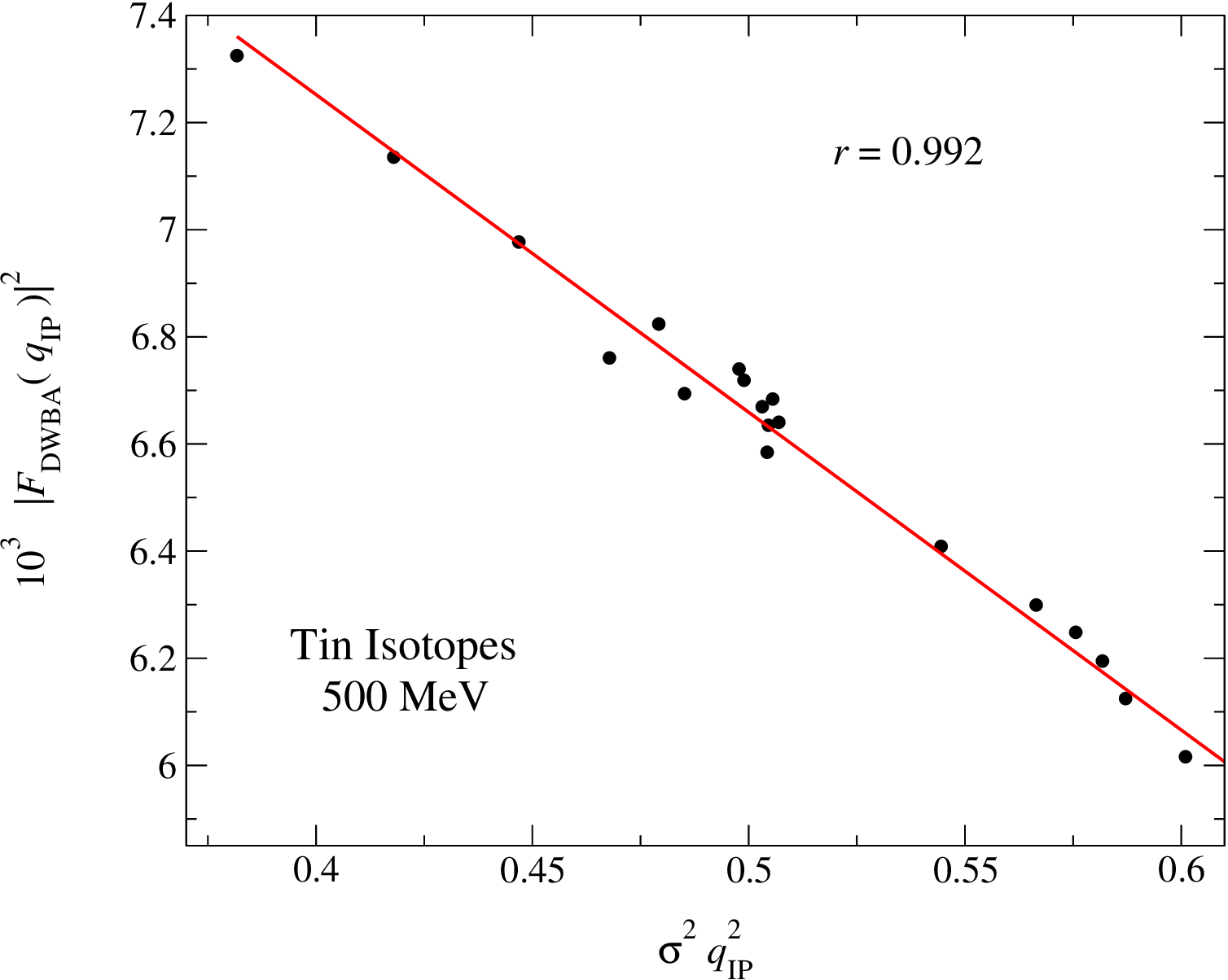}
\caption{\label{fsqtin} Correlation along the Sn isotopic chain of the
squared charge form factor in DWBA, at its first inflection point
($q_{\rm IP}$), with the product of the Helm parameter $\sigma^2$ and
$q_{\rm IP}^2$. The results are computed with the G2 effective interaction for
scattering by 500 MeV electrons. Note that each point of the figure
corresponds to a different isotope and that the mass number increases from
right to left.} 
\end{center}
\end{figure}

We now analyze the evolution along the isotopic chain of the momentum
transfer $q_{\rm IP}$ corresponding to the first inflection point of
$\vert F_{\rm DWBA}(q) \vert^2$ and the variation of $\vert F_{\rm
DWBA}(q_{\rm IP}) \vert^2$, calculated for an electron beam energy
of 500 MeV\@. We discuss possible correlations with the parameters
$R_{0}$ and $\sigma$ of the equivalent Helm density. Two noticeable
findings of this study are displayed in Fig.\ \ref{qartin}.
In the upper panel of the figure it is seen that the change of
the momentum transfer $q_{\rm IP}$ with mass number in the Sn isotopic
chain approximately follows an $A^{-1/3}$ law. Thus, the position of
the first minimum of the square of the modulus of the electric charge
form factor is shifted towards smaller $q$ values as the number of neutrons
in the isotopic chain increases, a feature also noted in Ref.\
\cite{Zai05}. Another interesting result is that the momentum transfer
$q_{\rm IP}$ turns out to be almost proportional to the effective
momentum transfer $q_{\rm eff}$ corresponding to the corrected first
minimum of  
$|F_{\rm PWBA}(q)|$ computed with the mean field charge density. The
situation is illustrated in the middle panel of Figure \ref{qartin}.
Recalling Eq.\ (\ref{eq18}), this correlation allows one to establish
a straightforward relationship between $q_{\rm IP}$ and the
Helm parameter $R_{0}$ in the tin isotopic chain:
\begin{equation}
R_0 \approx \frac{4.934}{q_{\rm IP}} .
\label{eq18bis}
\end{equation}
We also find that the rms charge radii of
the tin isotopes exhibit a considerable linear correlation with the
momentum transfer $q_{\rm IP}$, as depicted in the lower panel of
Fig.\ \ref{qartin}. These correlations could, in principle, provide
an alternative way to obtain the parameter $\sigma$ of the equivalent
Helm charge densities directly through Eq.\ (\ref{eq16}), taking into
account the relationship (\ref{eq18bis}) between $R_{0}$ and $q_{\rm
IP}$. 

Figure \ref{fsqtin} displays the evolution in the Sn chain of $|F_{\rm
DWBA}(q_{\rm IP})|^{2}$ (the value of the DWBA squared charge form
factor calculated at $q_{\rm IP}$) as a function of $\sigma^{2}q_{\rm
IP}^2$. The addition of neutrons along the isotopic chain in general
brings about an increase of the value of $|F_{\rm DWBA}(q_{\rm
IP})|^{2}$, also documented in the literature
\cite{Ant05,Sar07,Bertu07,Zai04,Zai05}. Furthermore, we notice that an
interesting linear correlation arises between the quantities $|F_{\rm
DWBA}(q_{\rm IP})|^{2}$ and $\sigma^{2}q_{\rm IP}^2$ as one moves from
the proton-rich side of the isotopic chain to the neutron-rich side.
This correlation may be qualitatively understood in the following
terms. If for guidance we consider the expression (\ref{eq17}) of the
electric charge form factor in the adopted Helm model, we see that the
natural variables to investigate the variation of the charge form
factor are $q R_0$ and $\sigma^2 q^2$. But, as stated in Eq.\
(\ref{eq18bis}), the value of $q R_0$ at the first inflection point of
$|F_{\rm DWBA}(q)|^{2}$ is practically independent of the mass number
in the whole isotopic chain. Therefore, the mass-number variation of
$\sigma^{2}q_{\rm IP}^2$ is left as the principal source for the
$A$-dependence of the value of the squared charge form factor at
$q=q_{\rm IP}$. It is then reasonable that the change with $A$ of
$|F_{\rm DWBA}(q_{\rm IP})|^{2}$ and $\sigma^{2}q_{\rm IP}^2$ is
correlated along the isotopic chain, a feature confirmed by Fig.\
\ref{fsqtin}. As a physical insight, helped by Eq.\ (\ref{eq18bis}),
the product $\sigma^2q_{\rm IP}^2$ can be recast as proportional to
$\sigma^2/R_0^2$, which is the ratio between the surface width and the
mean location of the surface of the underlying nuclear charge density.

\begin{figure}
\begin{center}
\includegraphics[width=0.95\linewidth,angle=0,clip=true]{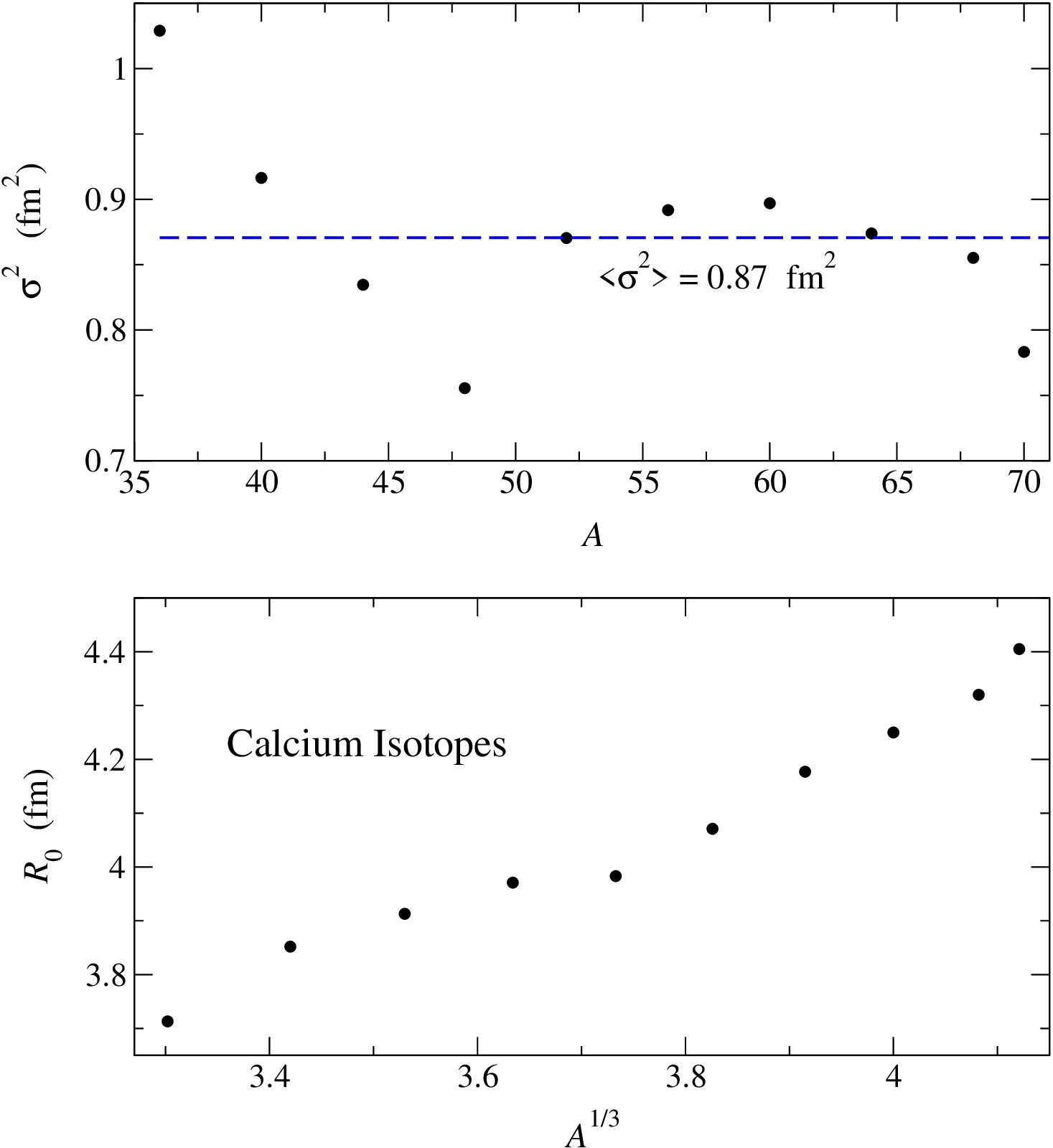}
\caption{\label{hparCa}
Same as Fig.\ \ref{hpartin} for the Ca isotopic chain.
}
\end{center}
\end{figure} 

\begin{figure}
\begin{center}
\includegraphics[width=0.98\linewidth,angle=0,clip=true]{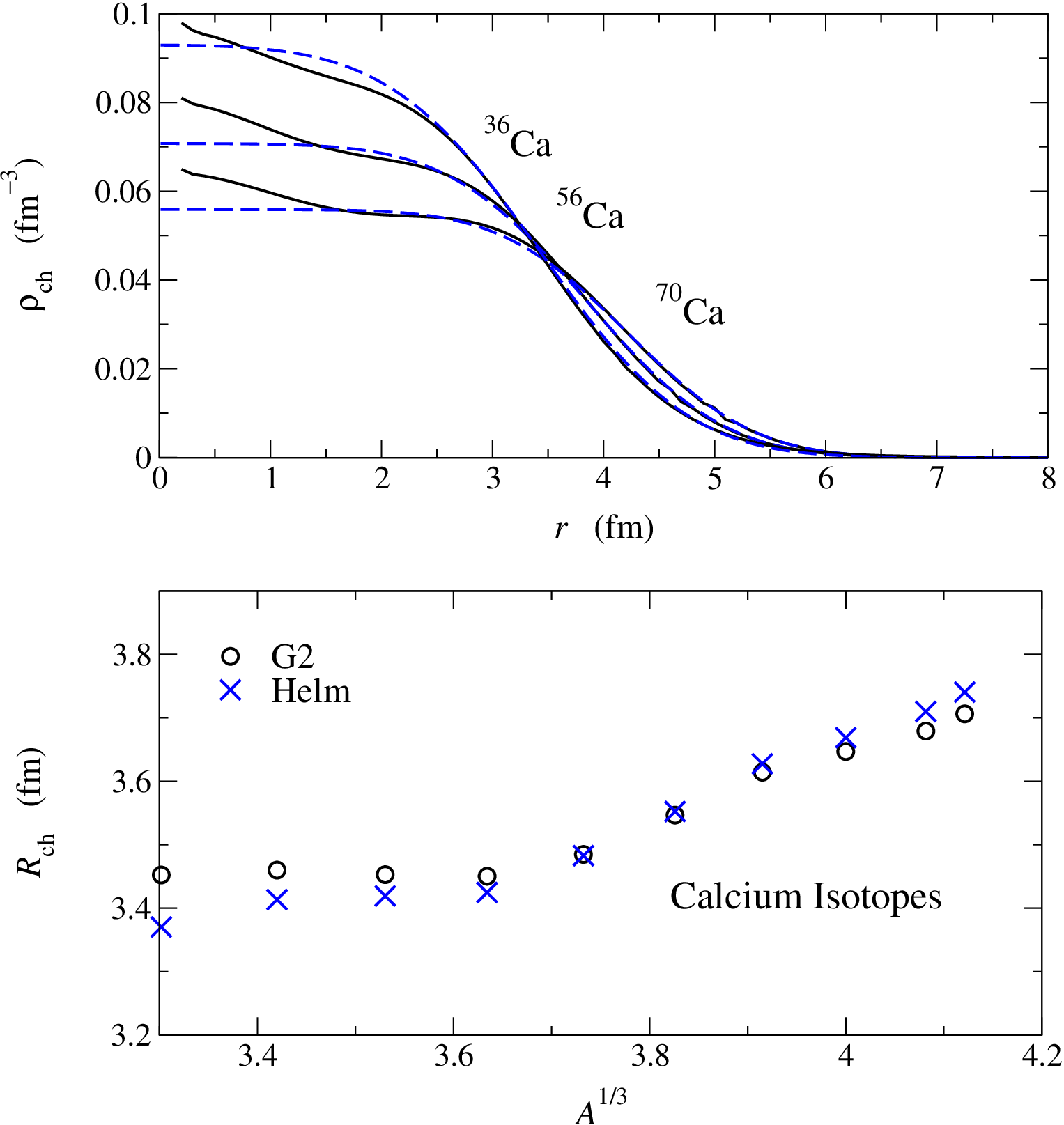}
\caption{\label{drchCa} 
Same as Fig.\ \ref{drchtin} for the Ca isotopic chain.
}
\end{center}
\end{figure}

\begin{figure}
\begin{center}
\includegraphics[width=0.95\linewidth,angle=0,clip=true]{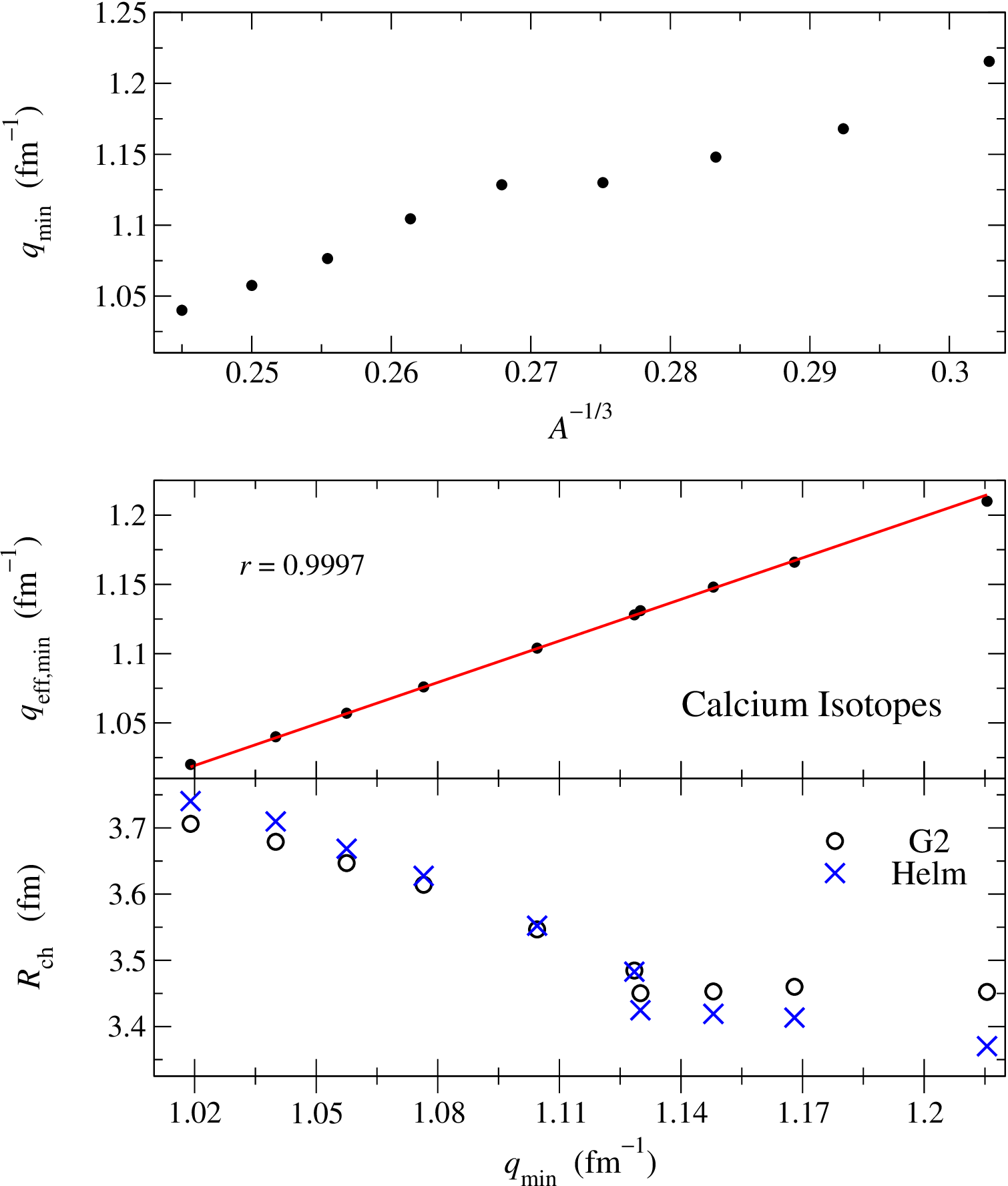}
\caption{\label{qarCa} 
Same as Fig.\ \ref{qartin} for the Ca isotopic chain.}
\end{center}
\end{figure}

\begin{figure}
\begin{center}
\includegraphics[width=0.95\linewidth,angle=0,clip=true]{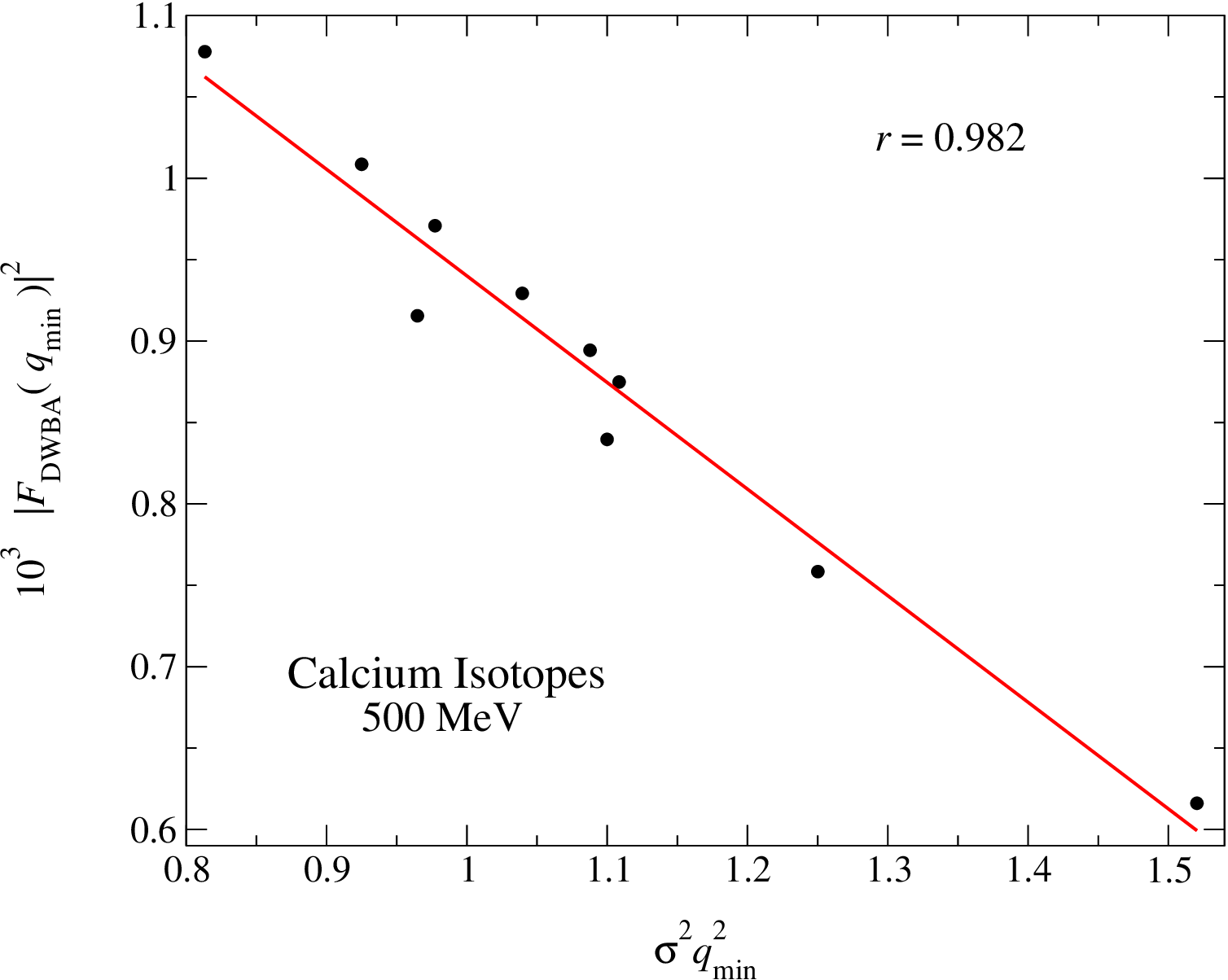}
\caption{\label{fsqCa} 
Same as Fig.\ \ref{fsqtin} for the Ca isotopic chain.}
\end{center}
\end{figure}

Similar results to those discussed above for tin have been obtained in
the study of the calcium isotopic chain. They are presented in Figures
\ref{hparCa}--\ref{fsqCa}. We first have modeled the calculated mean
field densities of the calcium chain by the simpler form of Helm
densities. The mass-number evolution of the parameters of the Helm
charge densities is illustrated in Figure \ref{hparCa}. The shrinkage
of the Helm parameter $\sigma$ at magic neutron numbers, noticed in
the tin chain, is a feature also present in the calcium isotopes for
the magic neutron numbers $N=28$ and $50$. However, the effect at
$N=20$ is completely washed out, in agreement with the result of Ref.\
\cite{Frie82}. The mean field charge densities of $^{36}$Ca, $^{56}$Ca, and
$^{70}$Ca calculated with G2, along with the equivalent Helm profiles, are 
displayed in the upper panel of Figure \ref{drchCa}. As in the case of
the tin chain, the effects of the addition of neutrons are very
manifest in the mean field charge density, as one can appreciate from
the changes in the interior and surface regions of the density
distributions of the shown Ca isotopes. One also sees that the
Helm densities manage to follow on average these changes. The Helm
profiles are found to reproduce closely the surface region of the
mean field densities, including the isotopes at the proton- and
neutron-rich sides of the calcium chain. Discrepancies are
observed in the interior region of the density distributions.
In the lower panel of Figure \ref{drchCa} we display, as a function of
$A^{1/3}$, the variation in the calcium chain of the rms radii of the
mean field and Helm charge densities. Compared to the case of Sn, 
this lighter chain presents more significant departures from the
$A^{1/3}$ behavior.  Also, the agreement of the rms radii of the Helm
model with the self-consistent G2 values is less good. This is more
visible in approaching the drip lines, especially at the proton drip
line.

Analogous correlations to those previously discussed in the case of
tin, between the value of $q$ at the first minimum (or IP) of $\vert
F_{\rm DWBA}(q) \vert^2$ with (i) the mass number, (ii) the first
minimum of $|F_{\rm PWBA}(q)|$, and (iii) the rms radius of the charge
distributions, are similarly found in the analysis of the calcium
isotopes. They are displayed in Figure \ref{qarCa}. In the present
case, however, a significant departure of the radii $R_0$ and $R_{\rm
ch}$ from the ${A}^{1/3}$ law is observed as one moves towards the
proton drip line. These deviations may be largely due to the fact
that, in approaching the proton drip line, the protons are more
loosely bound and therefore the charge density extends to larger
distances compared with the stable nuclei above ${}^{40}$Ca. The
effect is much more prominent in calcium than in tin because of the
lower Coulomb barrier due to its smaller atomic number. In turn, the
same effect may originate that the Helm parameter $\sigma$ of
${}^{40}$Ca, measuring the surface thickness of the nucleus, does not
decrease as compared with the heavier neighbor nuclei (see upper panel
of Figure \ref{hparCa}). We plot in Fig.\ \ref{fsqCa} the square of the
DWBA charge form factor computed for 500 MeV electrons at its first
minimum against the value of the product $\sigma^{2}q_{\rm min}^2$. In
calcium, as in the case of the tin isotopes, an outstandingly linear
correlation exists between both quantities.

\begin{figure}[t]
\begin{center}
\includegraphics[width=0.95\linewidth,angle=0]{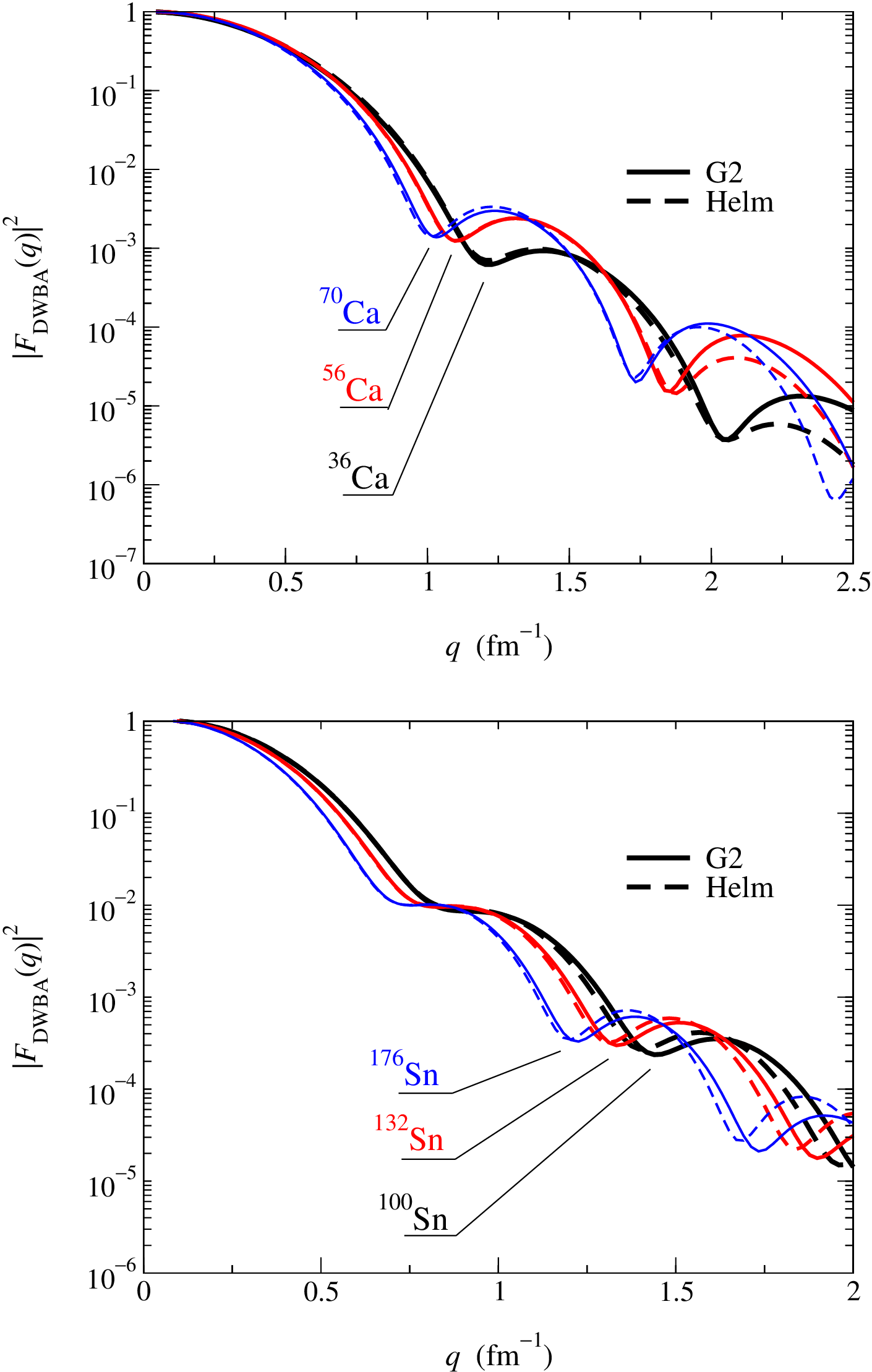}
\caption{\label{compff} 
(Color online) Comparison in Ca and Sn isotopes between the squared charge
form factor in DWBA calculated with the charge densities of the G2
mean field interaction (solid lines) and with the equivalent 
Helm charge densities (dashed lines). Notice that the vertical and
horizontal scales for calcium and tin are different.}
\end{center}
\end{figure}

Finally, in order to validate the consistency of our analysis with
Helm density equivalents, we compute in test cases the squared DWBA
electric charge form factor both with the Helm profiles and with their
original mean field charge densities. The results, as a function of
the momentum transfer $q$, are compared in Fig.\ \ref{compff} in a few
isotopes of calcium and tin. The cases shown in the figure are chosen to
illustrate the situation from the proton to the neutron drip lines, but
similar conclusions are found for the other isotopes of these chains. In the
range of $q$ values up to $\approx$1.5 fm$^{-1}$ in the calcium isotopes and
up to around $\approx$1 fm$^{-1}$ in the tin isotopes, one observes an 
excellent agreement between the results for $\vert F_{\rm DWBA}(q)
\vert^2$ obtained using the mean field charge densities and using the
equivalent Helm charge densities.
A similar situation is found in the nuclei that lie near the
drip lines: the agreement between the Helm and the mean field results
is slightly worse but without sizable
differences compared to the more stable nuclei. This scenario points
out that the electron-nucleus scattering results in the low-momentum
transfer region computed with mean field charge densities, are very
well simulated by equivalent Helm densities with the $R_0$ and
$\sigma$ parameters obtained according to the method described above.
As the value of the momentum transfer increases, expectably, the
discrepancies between the actual calculations with G2 and the adjusted
Helm model become evident.

Let us now concentrate on the results pertaining to the
self-consistent calculations with G2 that are shown in Fig.\
\ref{compff}. One can observe that, as we have mentioned earlier in the text,
the first minimum that is visible in the squared 
charge form factor of calcium sinks away in tin. Its fingerprint in
tin is recognized as an inflection point. The second and further
minima of $\vert F_{\rm DWBA}(q) \vert^2$ remain, however, clearly
identifiable in the tin isotopes. With the progressive addition of
neutrons in either of the two isotopic chains, it may be seen from
Fig.\ \ref{compff} that the first minimum of $\vert F_{\rm DWBA}(q)
\vert^2$, or its signature, becomes slightly more marked. This effect
is, however, much less noticeable than the discussed effect on the
first minimum induced by changing $Z$. One also observes that the
location of the minima or inflection points of the squared charge form
factor is gradually pushed towards lower momentum transfers as the
nucleus becomes more and more neutron rich. This effect has been noted
in previous literature \cite{Ant05,Sar07,Bertu07,Zai04,Zai05}, and for
the first minimum we have described it in more detail in the upper
panels of Figs.\ \ref{qartin} and \ref{qarCa}. The same trend (inwards
shift of the minima or IP) occurs with increasing atomic number
(compare the scales of the $q$ axis for Ca and Sn in Fig.\
\ref{compff}). In general, the inwards displacement of the momentum
transfer of the minima is accompanied by a simultaneous increase of
the height of the maxima of the squared charge form factor.

\section{Summary and conclusions}

We have adapted the {\sc elsepa} code \cite{Sal05} for
calculations of elastic electron-nucleus scattering. The predictions
obtained from the charge densities computed with various selected Skyrme
forces and modern relativistic mean field parameter sets have been 
compared with existing experimental data about elastic electron 
scattering in several stable nuclei.

A suitable quantitative comparison among the theoretical predictions 
is established by introducing, for each nucleus and beam
energy, a normalized square weighted difference. While all the
considered effective interactions describe qualitatively well the
experimental DCS data, the quantitative analysis in terms of the $d_{\rm w}^2$
with respect to experiment shows some differences among the various 
theoretical DCS, especially for large scattering angles.
These differences are related mainly with the different behavior 
of the mean field densities in the inner region of the nuclei. For the
investigated nuclei and energies, one always finds a theoretical
charge distribution whose $d_{\rm w}^2$ is at least similar, and in some cases
better, than the one obtained with the experimental charge density
that has been fitted to reproduce the measured DCS data. Among the
nuclear interactions analyzed, the DCS values calculated with the
charge densities obtained from the covariant DD-ME2 and FSUGold
parameter sets are the ones that tend to give overall better
agreement with the experimental values considered. 

A more challenging
test on theory is provided by the analysis of the experimental data
about relative differences of DCS between pairs of neighbor nuclei.
The calculations based on the mean field charge densities reasonably
follow the global trends shown up by the experimental measurements.
The quantitative analysis, however, as in the case of the previously
investigated DCS, highlights some deficiencies of the mean field
densities in their inner region.

We have used the mean field charge densities obtained with the
relativistic G2 parameterization, by the reasons discussed previously,
as the input baseline to study the elastic electron-nucleus scattering
along the tin and calcium isotopic chains. Calculations have been
performed from the proton to the neutron drip lines. Our aim has been
to extract general trends, according to current mean field theories of
nuclear structure, about the behavior that may be expected from real
electron-nucleus scattering experiments in exotic nuclei, in the low-momentum
transfer region. Such experiments are envisaged at FAIR \cite{Sim07} and the 
SCRIT \cite{Sud05,Wak08} project in the nearby future, using exotic
nuclei provided by radioactive isotope beams.
We have confined our study to medium and heavy mass isotopes where the
mean field approach can be applied.
Surely, the experimental program will investigate not only heavy
exotic nuclei but it will address also scattering from light exotic
nuclei. The theoretical investigation of these light nuclei,
however, demands a sophisticated microscopic treatment of
scattering to deal with the underlying shell model structure and
the possible occurrence of halos, which is beyond the methodology of
the mean field study of the present work.

First, we have computed for each isotope of the investigated chains
the squared electric charge form factor. It has been obtained as the
ratio between the DWBA DCS calculated with {\sc elsepa} and the Mott
DCS\@. We have checked that $\vert F_{\rm DWBA}(q) \vert^2$ defined in
this way is relatively independent of the energy of the electron beam
up to momentum transfers $q\approx 1$--1.5 fm$^{-1}$, even for
nuclei
as large as $^{208}$Pb. Second, we have fitted the mean field charge
densities by two-parameter Helm distributions. In doing so, we have
adjusted the effective momentum transfer correction to reproduce the
mean field rms charge radii, on average, along the isotopic chain. We
have made an {\em a posteriori} check that a DWBA calculation of the
elastic electron scattering using as input the equivalent Helm charge
density, is in excellent agreement with the results computed with the
original mean field charge density up to momentum transfers $q\approx
1$--1.5 fm$^{-1}$.

We have paid special attention to the value of the square of the DWBA
electric charge form factor at the momentum transfer where its first
minimum, in medium-mass nuclei, or its first inflection point, in
heavier nuclei, appears. We have studied how it evolves along the
isotopic chains of tin and calcium. Interesting linear correlations
between the value of the momentum transfer at the first minimum (or
IP) of $|F_{\rm DWBA}(q)|^{2}$ with the mass number of the isotopes of
the chain, with the effective momentum transfer at the first minimum
of $|F_{\rm PWBA}(q)|^2$, and with the rms radius of the charge
distribution, have been discussed. Also, a linear correlation
between $|F_{\rm DWBA}(q)|^{2}$ and $\sigma^{2}q^2$
computed at the first minimum or IP, where $\sigma$ is
the Helm parameter that accounts for the surface thickness 
of the nuclear density, has been found in the studied chains.

The analysis described in the present paper could potentially be useful 
for future electron-nucleus elastic scattering experiments. If the
experimental data are available for two or more isotopes of a given
chain, the aforementioned linear correlations would provide, for an
unknown nucleus of the chain, a hint on the value expected for the
square of the experimental electric charge form factor at its first
minimum, and for the momentum transfer where the latter occurs. The
parameters of the Helm charge density distribution of the unknown isotope
could be estimated by means of correlations such as those displayed in
Figs.\ \ref{qartin} and \ref{qarCa}, with the help of Eqs.\ (\ref{eq18})
and (\ref{eq16}). Also, if the value of the squared modulus of the
form factor is determined experimentally at its first minimum, the
charge density in the Helm model can be sketched from similar
correlations to
Figs.\ \ref{qartin} and \ref{qarCa}, together with the correlation of
the type depicted in Figs.\ \ref{fsqtin} and \ref{fsqCa}. The use of
more elaborated versions of the Helm model \cite{Frie82, Frie86,
Spru92} that take into account the central depression of the charge
density, should allow one to extend the domain of validity of our
method up to larger values of the momentum transfer. Work in this
direction will be undertaken.

\acknowledgments
We are very grateful to Prof.\ P. Ring for supplying us with the code
for finite nuclei calculations with the model DD-ME2. We thank Prof.\
D.W.L. Sprung for reading a preliminary version of the manuscript and
for valuable comments. This work has been partially supported by
Grants Nos.\ FIS2005-03142 and FPA2006-12066 from MEC (Spain) and
FEDER, and by Grant No.\ 2005SGR-00343 from Generalitat de Catalunya,
as well as by the Spanish Consolider-Ingenio 2010 Programme CPAN
CSD2007-00042. One of us (X.R.) also acknowledges Grant No.\
AP2005-4751 from MEC (Spain).


%
\end{document}